\documentclass[twocolumn]{aastex631}
\usepackage[T1]{fontenc}
\usepackage[utf8]{inputenc}
\usepackage{lmodern}
\usepackage{amsmath,amsfonts,amssymb}
\usepackage{graphicx,xcolor}
\usepackage{esint}
\usepackage{nicefrac}
\usepackage{mathtools}
\usepackage{marginnote}
\DeclareMathAlphabet{\mathbfsf}{\encodingdefault}{\sfdefault}{bx}{sl}

\newcommand{\I}{\mathrm{i}}
\newcommand{\e}[1]{\mathrm{e}^{#1}}

\shorttitle{Numerical Nuances of Global Coronal Models}
\shortauthors{Brchnelova et al.}
\graphicspath{{./}{Figures/old/}{Figures/new/}}

\begin{document}

\title{To $\vec{E}$ or not to $\vec{E}$:\\ Numerical Nuances of Global Coronal Models}

\author[0000-0003-0874-2669]{Michaela Brchnelova}
\affiliation{Centre for mathematical Plasma Astrophysics, \\
KU Leuven, 3001 Leuven, Belgium}

\author[0000-0001-9438-9333]{B{\l}a{\.{z}}ej Ku{\'{z}}ma}
\affiliation{Centre for mathematical Plasma Astrophysics, \\
KU Leuven, 3001 Leuven, Belgium}

\author[0000-0002-2137-2896]{Barbara Perri}
\affiliation{Centre for mathematical Plasma Astrophysics, \\
KU Leuven, 3001 Leuven, Belgium}

\author[0000-0003-4017-215X]{Andrea Lani}
\affiliation{Centre for mathematical Plasma Astrophysics, \\
KU Leuven, 3001 Leuven, Belgium}

\author[0000-0002-1743-0651]{Stefaan Poedts}
\affiliation{Centre for mathematical Plasma Astrophysics, \\
KU Leuven, 3001 Leuven, Belgium}
\affiliation{Institute of Physics, University of Maria Curie-Sk{\l}odowska,\\
Pl.\ M.\ Curie-Sk{\l}odowskiej 5, 20-031 Lublin, Poland}


\correspondingauthor{Michaela Brchnelova}
\email{michaela.brchnelova@kuleuven.be}

\begin{abstract}
In the recent years, global coronal models have experienced an ongoing increase in popularity as tools for forecasting solar weather. Within the domain of up to $21.5\;R_\odot$, magnetohydrodynamics (MHD) is used to resolve the coronal structure using magnetograms as inputs at the solar surface. Ideally, these computations would be repeated with every update of the solar magnetogram so that they could be used in the ESA Modelling and Data Analysis Working Group (MADAWG) magnetic connectivity tool (\url{http://connect-tool.irap.omp.eu/}). Thus, it is crucial that these results are both accurate and efficient. While much work has been published showing the results of these models in comparison with observations, not many of it discusses the intricate numerical adjustments required to achieve these results. These range from details of boundary condition formulations to adjustments as large as enforcing parallelism between the magnetic field and velocity. By omitting the electric field in ideal-MHD, the description of the physics can be insufficient and may lead to excessive diffusion and incorrect profiles. We formulate inner boundary conditions which, along with other techniques, reduce artificial electric field generation. Moreover, we investigate how different outer boundary condition formulations and grid design affect the results and convergence, with special focus on the density and the radial component of the $\mathbf{B}$-field. The significant improvement in accuracy of real magnetic map-driven simulations is illustrated for an example of the 2008 eclipse.

\end{abstract}




\section{Introduction}

Global coronal models based on CFD (Computational Fluid Dynamics) methods applied to MHD equations are becoming \textcolor{black}{a widely used} 
tool to aid solar weather predictions \textcolor{black}{\citep[see e.g.,][and reference therein]{Usmanov1996,Linker1999,vanderHolst2014,Pinto2017,Mikic2018,Reville2020}. In these models}, solar photospheric  magnetic map data are used to prescribe the radial component of the magnetic field on the inner boundary of the domain, representing the lower corona and then, MHD equations are solved in the rest of the domain. This domain typically extends over roughly 21 solar radii, beyond which faster and more idealised heliospheric codes, such as EUHFORIA 
\textcolor{black}{\citep[a code for space weather predictions of the solar wind and CME evolution, see, e.g.,][]{Poedts,Pomoell2018,Scolini2018,Scolini2019}}, can be coupled to study the propagation of the coronal features to the rest of the solar system.

Since in this case, the outer boundary data resulting from the MHD simulation are used as an input for EUHFORIA (see an example of such application in \cite{Samara2021}) or other codes to detect possible geomagnetic storms, it is crucial that the features (e.g.\ coronal streamers) in the MHD solutions are accurately resolved. \textcolor{black}{To this end, the boundary conditions must be formulated appropriately and the numerical phenomena arising in the simulations understood possibly avoided, which is the aim of this paper.
One such problematic numerical phenomenon arises from the fact that the ideal (single-fluid) MHD equations are only a simplified version of a more detailed multi-fluid description of solar plasmas}. Unlike multi-fluid models, they solve for the electric field only implicitly, without resolving it in absolute terms. It is shown in this paper that not properly accounting for the electric field in ideal MHD when formulating the inner boundary condition leads to a loss of accuracy and sharpness in the results and, in the worst case scenario, even leads to completely incorrect profiles.

Determining realistic values of the electric field in the solar corona is problematic. While there exist techniques to estimate the photospheric electric field such as inversion from the observed $\mathbf{V}$- and $\mathbf{B}$-fields (see for example the work of \cite{Lumme2016}), getting the estimate of the electric field in the corona is not possible in this way as we do not have direct observations of the magnetic field. The magnitude of the electric field in the quiet corona is typically expected to be very small owing to the conductivity of the solar wind plasma \citep{Kislov2022}, \textcolor{black}{at least outside of regions of large velocities where even a small misalignment between the velocity and the magnetic field could cause large $\mathbf{E}$-fields}. For that reason, also magnetostatic solutions, in which the electric fields relax to zero, are used to represent and predict the coronal behaviour \citep{Contopoulos2011}. Locally however, the electric field can be enhanced by, for example, reconnection events in the current sheets or coronal holes, see the work of \cite{Zank2014} or \cite{Zharkova2015}. Parallel electric field generation due to Alfvén or fast-magnetosonic waves is also predicted by for example \cite{Kaghashvili2012}, in which case the predicted magnitudes range from 1$\times$10$^{-2}$ to 1$\times$10$^{-1}\;$mVm$^{-1}$.

\textcolor{black}{In our setup, we are looking at a steady-state background corona, where we would expect the resulting electric field should to be very small. This is because modelling of the above-mentioned $\mathbf{E}$-field enhancing effects cannot be done properly for two reasons; one being that our simulation is not time-dependent to model waves and second, that the majority of the reconnection events observed in our simulations are driven by excessive numerical diffusion instead of being physical, as will be shown later in the paper.}

To obtain sharp features and ensure a physical alignment of the $\mathbf{V}$- and $\mathbf{B}$-fields, the parallelism can be enforced artificially. This can be done by, for example, aligning the resulting $\mathbf{B}$-field with the velocity starting from a certain distance from the Sun. A second option, recently proposed by \cite{sokolov2021streamaligned}, consists of solving reduced ideal MHD equations. However, none of these two approaches may be easy to implement into an existing  numerical code, because the former requires to override the $\mathbf{B}$-field solutions during computation (which is far from trivial in, e.g., time-implicit solvers like ours in which both the system Jacobian matrix and the right-hand-side vector would need to be manipulated in complex ways) and the latter requires to completely alter the ideal MHD formulation. 
Some other techniques can be used, however, to reduce the artificial $\mathbf{E}$-field generation and thus improve the sharpness and accuracy of the features. It will be shown that, for instance, one of the most crucial elements is the formulation of the inner boundary condition (BC). 

Even though the outer BC is supersonic and hence, in theory, should not have much effect on the solution in the domain, we will show that this is not true. Most authors, when referring to the outer BCs of coronal models, mention that the state values of the boundary cells are "extrapolated" to the ghost cells (see, for example, \cite{Pomoell2011}). However, the manner in which this extrapolation is performed affects the solution significantly. As will be shown in this paper, aspects such as the orientation and topology of the $\mathbf{B}$-field lines of streamers can be directly manipulated depending on \textit{how} the magnetic field is extrapolated into the ghost cells. 

Finally, many of these undesired effects can be reduced by a careful grid design (see \cite{Brchnelova}), which is another crucial aspect that will be discussed. This grid type limits the extent to which the outer BC can affect the domain and the coupling to heliospheric codes such as EUHFORIA. All of these above mentioned aspects are discussed in this paper.

\textcolor{black}{This paper is organised as follows. 
In Section~\ref{sec:theory} better insight into the origins of the $\mathbf{E}$-field problem is provided through the comparison of the full multi-fluid MHD and the ideal MHD equations. For completeness' sake, the solver used in this study is also presented. In Section~\ref{sec:mechanisms}, the mechanisms of the artificial $\mathbf{E}$-field generation in our simulations are identified, and their effects on the solution demonstrated on a steady-state dipole case. It is shown that this mostly relates to the electric-field production on the inner boundary, and thus this effort leads to outlining recommendations on the formulation of the inner BC. With the inner BC discussed, in Section \ref{sec:outer}, the outer BC is addressed, where the effects of various extrapolation laws are analysed. It is also shown that it is not possible to prescribe a universally correct law for the entire domain, leading to the introduction of a new type of grid that can help reduce the outer-BC-related effects in problematic regions. Finally, in Section~\ref{sec:map}, the proposed techniques are applied to a data-driven simulation of the global corona during the 2008 solar eclipse. The last Section concludes our work and summarises recommendations for procedures that could be standardised for global coronal models in the future.
}

\section{Theoretical context}
\label{sec:theory}

\textcolor{black}{Before diving into the design of boundary conditions, we should first better understand the problem at hand. First, the solver we work with in this paper is introduced and afterwards, the origin of the  $\mathbf{E}$-field issue is illustrated on comparing the full multi-fluid model equations with the ideal MHD equations.}

\subsection{Introduction to the COCONUT solver}
\label{subsec:CF}

The ideal MHD solver used in this paper is COCONUT; a solver based on the COOLFluiD (Computational Object-Oriented Libraries for Fluid Dynamics) platform. The latter has been developed since many years for scientific multi-physics simulations \citep{Lani2005, Kimpe2005, Lani2013}, including high-speed flows, radiation, thermo-chemical nonequilibrium as well as plasma flows \textcolor{black}{\citep[see e.g.,][and reference therein]{Santos2016, Lani2011, LaniGPU, Asensio2019}.} 

The details of the ideal MHD model for global coronal modelling, its verification and initial validation are presented in \cite{COCONUT}, where more details about the numerical set-up can be found. Unlike the majority of global coronal models, this solver relies upon an implicit scheme, i.e.\ the backward Euler time discretization scheme, for steady state simulations. This allows it to converge much faster compared to explicit schemes, as the Courant-Friedrichs-Lewy (CFL) number can largely exceed the value of 1 (up to 1000 or more depending on the test case) during the computation. This feature makes it a promising tool for real-time data-driven solar simulations. The solver also works with unstructured grids, which allows us to experiment with grid resolutions and topologies to optimise the solver performance (see \cite{Brchnelova}). In the future, it is also planned to integrate adaptive mesh refinement (AMR) \citep{BenAmeur2021} and high-order Flux Reconstruction algorithms \citep{Vandenhoeck2019} into this model, which would dramatically increase the accuracy (potentially up to $10^{th}$ order) on coarser grids and, possibly, accelerate the convergence of the solver even further.

The code uses a second-order accurate finite volume (FV) discretization along with the hyperbolic divergence cleaning (HDC) to ensure that the divergence of the $\mathbf{B}$-field stays zero. The MHD equations are solved on arbitrarily unstructured grids in a nondimensional form. The full form can be found in \cite{COCONUT} along with the formulation of the baseline BCs. In this study however, the rotation is not considered. The reference values are set to:
\begin{equation}
    \rho_\text{ref} = 1.67 \cdot 10^{-13}\;\text{kg.m$^{-3}$}, \quad B_\text{ref} = 2.2 \cdot 10^{-4}\;\text{T},
\end{equation}
\noindent with then $V_\text{ref}$ computed to be the Alfvén velocity ($4.5 \cdot 10^{5}\;\text{m.s$^{-1}$}$), while the length-scales and coordinates are adimensionalised according to the solar radius.  

\begin{figure*}[t!]
\centering
\gridline{\fig{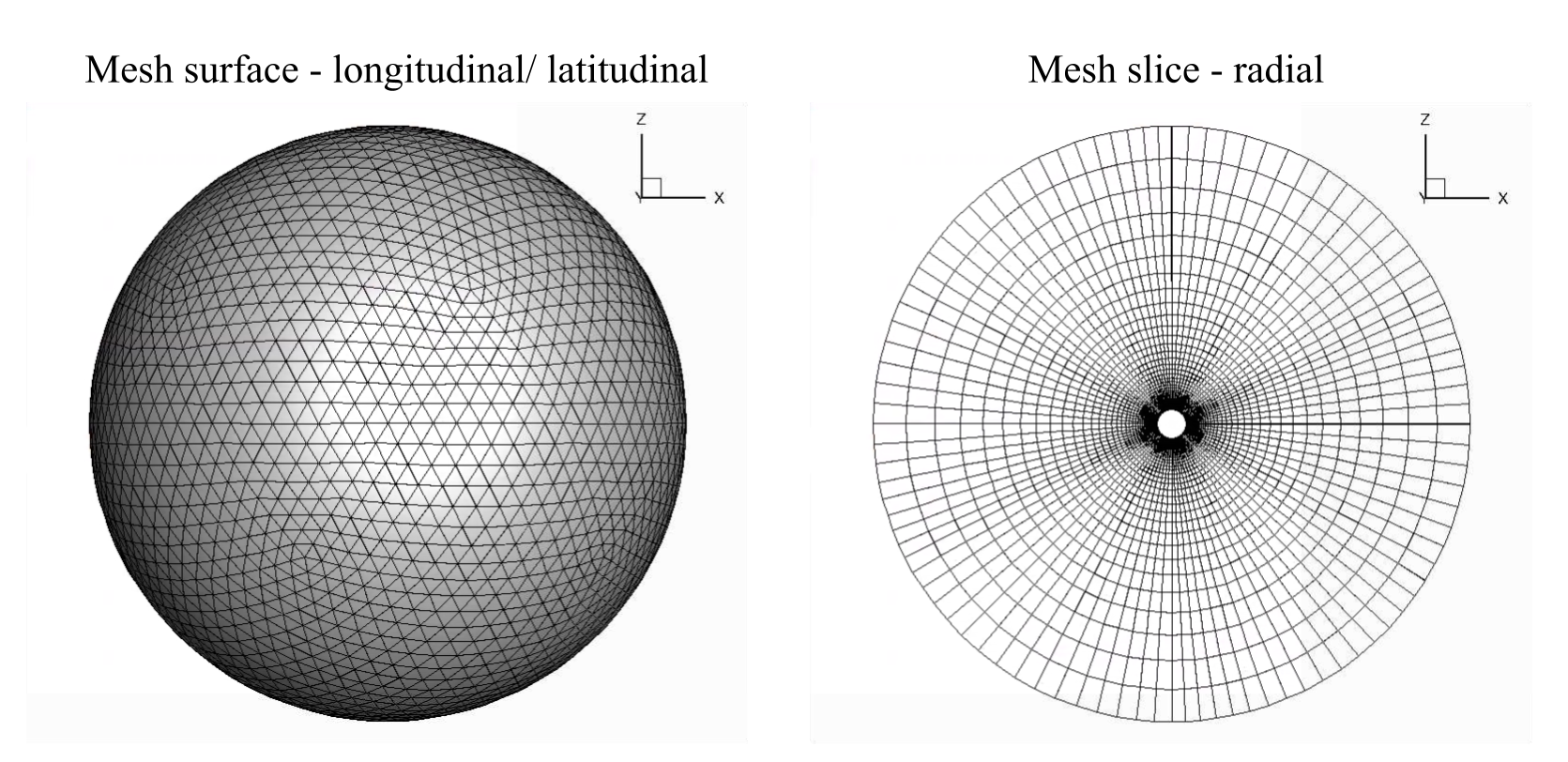}{0.9\textwidth}{}}
\caption{The mesh used for most simulations in this paper is based on prisms and has 300k elements. We show a 3D rendering of the outer spherical slice at 21 $R_{Sun}$ on the left panel, and a 2D vertical cut to focus on the stretched radial structure in the right panel.}
\label{fig:Mesh_details}
\end{figure*}

All of the simulations in this report, unless otherwise specified, were performed without the use of a limiter and with the 5-th subdivision level icosahedron-based spherical mesh with prisms (see \citet{Brchnelova} for details), with roughly 300k elements. The surface of the spherical shell domain, starting from 1.1$R_\odot$ to 21.5$\;R_\odot$, is shown in Figure~\ref{fig:Mesh_details} along with a cut-through showing the radial discretisation. When other grids are used in the paper (e.g., the grid designed to minimise outer-BC effects introduced later), it will be mentioned explicitly. 

Finally, before proceeding to the physics, convergence of the solver should be touched upon. When discussing residual levels in the text, the following definition of a residual is used:

\begin{equation}
    \mathrm{res}(a) = \log\sqrt{\sum_i\left(a_i^t - a_i^{t+1}\right)^2},
\end{equation}
\noindent in which $a$ is the relevant physical quantity, $i$ the spatial index and $t$ the temporal index during the pseudo-time stepping. In general, the velocity and the magnetic field components are monitored during convergence. The simulation is typically considered converged when the velocity residual drops below -3 to -4. For comparison analyses in this paper, however, the residual was brought down to -10 in $V_x$ to ensure that the differences observed in the flow field are indeed due to the investigated aspects (e.g., the BCs) and not due to insufficient convergence. \textcolor{black}{It could be argued that looking at the squares of absolute value differences might bias our focus to regions where these values are by default higher, e.g. the outer boundaries when it comes to the velocity field. For this reason however, we also monitor the other residuals, like the pressure, density and the magnetic field, where the higher values are generally near the inner boundary instead. Via ensuring that all of the residuals are sufficiently small, the evaluation of convergence is more or less balanced over the domain. }

Now that the basic numerics has been discussed, the physics can be elaborated on in further details. 

\subsection{The origin of the $\mathbf{E}$-field problem}

\textcolor{black}{To outline the role of the electric field and why it should be addressed in the inner BC design,} it is instructive to show the full MHD equations considering both ion (subscript "$i$") and electron (subscript "$e$") fluids separately, as they both contribute to the generation and removal of the electric field owing to their charges. The resulting two-fluid model, in a dimensional form, can be formulated as follows \citep[see e.g.,][]{Meier2012}. The (pre-)Maxwell equations read:

\begin{equation}
    \mathbf{\nabla} \times \mathbf{E}=-\frac{\partial \mathbf{B}}{\partial t}, \quad \mathbf{\nabla} \times \mathbf{B}= \mu_{0} \mathbf{J},
\label{eq:Maxwell}
\end{equation}

\begin{equation}
    \mathbf{\nabla} \cdot \mathbf{B}=0, \quad  \mathbf{\nabla} \cdot \mathbf{E}=\sigma / \epsilon_{0},
\label{eq:Maxwell2}
\end{equation}

\noindent where $\sigma$ is the charge density ($\sigma = n_e q_e + n_i q_i$) and $\mathbf{J}$ the electric current density ($\mathbf{J} = n_i q_i \mathbf{V}_i + n_e q_e \mathbf{V}_e$). For our two species, here ions and electrons, the continuity equations can be written as:

\begin{equation}
\frac{\partial n_i}{\partial t}+\mathbf{\nabla} \cdot\left(n_i \mathbf{V}_{i}\right)= \\ 0 \quad \text{and} \quad \frac{\partial n_e}{\partial t}+\mathbf{\nabla} \cdot\left(n_e \mathbf{V}_{e}\right)=0,
\end{equation}

\noindent respectively. The momentum conservation is, for electrons:


\begin{multline}
\frac{\partial}{\partial t}\left(m_{e} n_e \mathbf{V}_{e}\right)+\mathbf{\nabla} \cdot\left(m_{e} n_e \mathbf{V}_{e} \mathbf{V}_{e}+{P}_{e}\right)= \\ -e n_e\left(\mathbf{E}+\mathbf{V}_{e} \times \mathbf{B}\right)+m_{e} n_e \mathbf{g} + \mathbf{R}_{i}^{i e},
\end{multline}

\noindent and for ions: 

\begin{multline}
\frac{\partial}{\partial t}\left(m_{i} n_i \mathbf{V}_{i}\right)+\mathbf{\nabla} \cdot\left(m_{i} n_i \mathbf{V}_{i} \mathbf{V}_{i}+{P}_{i}\right)= \\ e n_i\left(\mathbf{E}+\mathbf{V}_{i} \times \mathbf{B}\right)+m_{i} n_i \mathbf{g} +\mathbf{R}_{e}^{e i},
\end{multline}

\noindent where $P$ is the pressure, $\mathbf{g}$ the gravitational acceleration and $\mathbf{R}_{i}^{i e}$ and  $\mathbf{R}_{e}^{e i}$ the collisional momentum terms. If the collisional frequency between ions and electrons can be approximated through:

\begin{equation}
\nu_{i e} \sim \frac{\nu_{e i}}{N_{c}} \sim \frac{m_{e}}{m_{i}} \nu_{e i} \sim \frac{Z^{2} e^{4} n_{i} m_{e}^{1 / 2} \ln \Lambda_{e i}}{\left(4 \pi \epsilon_{0}\right)^{2} m_{i} T_{e}^{3 / 2}},
\end{equation}

\noindent with $\ln \Lambda$ being the Coulomb logarithm and $Z$ of 1, then the momentum collisional exchange terms can be computed through:

\begin{equation}
    \mathbf{R}_{i}^{i e} = \frac{m_i m_e}{m_i + m_e} n_i \nu_{i e} (\mathbf{V_e} - \mathbf{V_i}) \quad \text{and} \quad \mathbf{R}_{i}^{i e}=-\mathbf{R}_{e}^{e i}.
\end{equation}

\noindent Finally, the energy conservation reads:

\begin{multline}
\frac{\partial \Psi_{e}}{\partial t}+\mathbf{\nabla} \cdot\left(\Psi_{e} \mathbf{V}_{e}+\mathbf{V}_{e} \cdot {P}_{e}+\mathbf{h}_{e}\right) =\\\mathbf{V}_{e} \cdot\left(-e n_{e} \mathbf{E}+m_{e} n_{e} \mathbf{g} +  \mathbf{R}_{e}^{e i}\right) + Q_{e}^{e i}, \quad 
\label{eq:MFMHDE1}
\end{multline}
and
\begin{multline}
\frac{\partial \Psi_{i}}{\partial t}+\mathbf{\nabla} \cdot\left(\Psi_{i} \mathbf{V}_{i}+\mathbf{V}_{i} \cdot {P}_{i}+\mathbf{h}_{i}\right) =\\\mathbf{V}_{i} \cdot\left(e n_{i} \mathbf{E}+m_{i} n_{i} \mathbf{g} +  \mathbf{R}_{i}^{i e} \right)+ Q_{i}^{i e}.
\label{eq:MFMHDE2}
\end{multline}

\noindent in which $\Psi_e$ and $\Psi_i$ are the total internal energies of the electrons and the ions, and $Q_{e}^{e i}$ and $Q_{i}^{i e}$ are the collisional work terms between the two species:

\begin{equation}
   Q_{i}^{i e} = \frac{1}{2} \Big( \mathbf{R}_{i}^{i e} \cdot (\mathbf{V_e} - \mathbf{V_i})  \Big) + 3 n_i \nu_{i e} k_B (T_e - T_i), \quad 
\end{equation}
and
\begin{equation}
   Q_{e}^{e i} = \frac{1}{2} \Big( \mathbf{R}_{e}^{e i} \cdot (\mathbf{V_i} - \mathbf{V_e})  \Big) + 3 n_i \nu_{e i} k_B (T_i - T_e).
\end{equation}

In the model presented above, the electric field $\mathbf{E}$ is typically one of the primitive variables that is solved for. In Equations~(\ref{eq:MFMHDE1}) and (\ref{eq:MFMHDE2}), we can see, for example, the tendency of the two-fluid MHD to respond to existing electric fields via their dynamics, as the two fluids react to the present $\mathbf{E}$-field in opposite ways. Charge separation, which can result from this response is, by itself, a mechanism to create an opposing electric field, see Equation~(\ref{eq:Maxwell}) and (\ref{eq:Maxwell2}). Looking at Equations~(\ref{eq:MFMHDE1}) and (\ref{eq:MFMHDE2}), the $\mathbf{E}$-field can be also a major contributor to the fluids' energy budgets if it has a non-zero value and if it is not perpendicular to the fluids velocity field. 

However, the form of the MHD equations that we generally use in global coronal models is simplified through a variety of assumptions to the single-fluid ideal MHD equations. In particular, assuming non-relativistic flows the displacement current and the electrostatic force are ignored in MHD. As a result, the electric field and the current density adopt a secondary role as they can be eliminated from the equations, i.e.\ they can simply be computed from the other variables without solving a differential equation ($\mathbf{E}=-\mathbf{V}\times\mathbf{B}$ and $\mathbf{J}=\mathbf{\nabla}\times\mathbf{B}/\mu_0$). The MHD equations then read (here in a non-dimensional form, as also used in our code):

\begin{equation}
    \frac{d\rho}{dt} + \mathbf{\nabla} \cdot (\rho \mathbf{V}) = 0,
\end{equation}

\begin{equation}
    \frac{d(\rho \mathbf{V})}{dt} + \mathbf{\nabla} \cdot \Big( \rho \mathbf{V} \otimes \mathbf{V} + \mathbf{I} \left( P + \frac{\mathbf{B}^2}{8 \pi} \right) - \frac{\mathbf{B} \otimes \mathbf{B} }{4\pi} \Big) = \rho \mathbf{g},
\end{equation}

\begin{equation}
    \frac{d\mathbf{B}}{dt}  + \mathbf{\nabla} \times \left( - \mathbf{V} \times \mathbf{B} \right) = \mathbf{0},
\label{eq:E10}
\end{equation}


\begin{multline}
     \frac{d}{dt}  \left( \rho \frac{\mathbf{V}^2}{2} + \rho \mathcal E + \frac{\mathbf{B}^2}{8 \pi}\right) + \\ \mathbf{\nabla} \cdot \left[ \left( \rho \frac{\mathbf{V}^2}{2} + \rho \mathcal E + P \right) \mathbf{V} - \frac{1}{4 \pi} (\mathbf{V} \times \mathbf{B}) \times \mathbf{B} \right] = \rho \mathbf{g} \cdot \mathbf{V}.
\label{eq:E20}
\end{multline}

\noindent where $\mathcal E$ is the non-dimensional internal energy and $\mathbf{I}$ the identity diadic. 

Assuming plasma with negligible resistivity, using the definition of $\mathbf{E}$ as being the cross product of the magnetic and velocity fields, Equation~(\ref{eq:E10}) and Equation~(\ref{eq:E20}) of ideal MHD can be turned into:

\begin{equation}
   \frac{d\mathbf{B}}{dt}  + \mathbf{\nabla} \times \left(\mathbf{E} \right) = \mathbf{0} \quad 
\label{eq:E1}
\end{equation}
and

\begin{multline}
     \frac{d}{dt}  \left( \rho \frac{\mathbf{V}^2}{2} + \rho \mathcal E + \frac{\mathbf{B}^2}{8 \pi}\right) + \\ \mathbf{\nabla} \cdot \left[ \left( \rho \frac{\mathbf{V}^2}{2} + \rho \mathcal E + P \right) \mathbf{V} -\frac{1}{4 \pi} (-(\mathbf{E})) \times \mathbf{B} \right] = \rho \mathbf{g} \cdot \mathbf{V}.
\label{eq:E2}
\end{multline}

Equations~(\ref{eq:E1}) and (\ref{eq:E2}) show the implicit inclusion of the $\mathbf{E}$-field in ideal-MHD. While in this sense, the $\mathbf{E}$-field is included in the model, these MHD equations do not account for its absolute value, but only its curl and the divergence of its cross product with the $\mathbf{B}$-field. \textcolor{black}{Unlike the full multi-fluid model, in the ideal MHD equations without resistivity, there are no mechanisms included which would set constraints on the \textit{amplitude} of the electric field in case it is accidentally artificially generated in the simulation during convergence, as already pointed out by \cite{sokolov2021streamaligned}.}

Starting with zero $\mathbf{E}$-field at the beginning of the simulation, the $\mathbf{E}$-field during the run will relax such that Equations~(\ref{eq:E1}) and (\ref{eq:E2}) are met. The contribution to the $\mathbf{E}$-field will thus come from its curl and from the divergence of its cross product with the $\mathbf{B}$-field. In principle, in steady cases, the former should not contribute to the rise of the $\mathbf{E}$-field since $\partial \mathbf{B}/\partial t$ is zero. Equivalently, if the $\mathbf{E}$-field was non-existent and remained non-existent during convergence, the $(-(\mathbf{E})) \times \mathbf{B}$ term would also be zero, regardless of the $\mathbf{B}$-field orientation. This would remove the contribution of this term to the energy budget (and thus possibly minimising the system's energy). 

However, a steady magnetic field is not something that time-stepping codes compute with, even for steady cases. The magnetic field starts at some distribution given by the initial state and keeps adjusting and changing during the time-stepping while converging to the steady-state, giving rise to $\partial\mathbf{B}/\partial t$ in between the iteration steps. This means that also the electric field will attain a certain distribution to satisfy the changing magnetic field. With the ideal-MHD formulation, as shown above, the curl of this resulting $\mathbf{E}$-field will be close to zero to satisfy Equation~(\ref{eq:E1}), and its orientation with respect to the $\mathbf{B}$-field will be such that Equation~(\ref{eq:E2}) is met, but there are no constraints on its actual value. 

\textcolor{black}{It should be emphasised that since the $\mathbf{E}$-field in this model represents the misalignment between the magnetic field and the velocity field, its value affects how well the plasma flow will follow the resolved magnetic field lines}. Since the $\mathbf{V}$ $\times$ $\mathbf{B}$ product enters the energy equation, ultimately, this does not only directly affect the $\mathbf{B}$- and $\mathbf{V}$-fields, but, as will be shown later, also other hydrodynamic variables such as temperature.

\section{Formulating the Inner Boundary Conditions through $\mathbf{E}$-field Analysis}
\label{sec:mechanisms}

Now that the $\mathbf{E}$-field issue was introduced, first, the sources of the spurious $\mathbf{E}$-field in the simulations are described. Next, the inner BC is re-formulated in order to reduce the amount of this spurious $\mathbf{E}$-field. Results are shown for the steady dipole for both large $\mathbf{E}$-field and low $\mathbf{E}$-field cases, through the comparison of which the effects of the $\mathbf{E}$-field and the impact of the new BC can be studied. Comparison of the results will point to another issue - which is the effect of numerical diffusion causing reconnection, which will be briefly discussed afterwards. 

\subsection{Mechanisms of $\mathbf{E}$-field generation}

Let us first look at the major sources of the $\mathbf{E}$-field in the simulations. Since the domain here investigated is spherical, the analysis from here on will be interpreted using the spherical coordinate system, $r, \theta$ and $\varphi$. During our analysis, the main possible sources were identified as follows:

\begin{itemize}
    \item the initial conditions;
    \item the inner and outer BCs;
    \item internal physical processes such as reconnection;
    \item the mesh topology and domain discretisation.
\end{itemize}

\noindent The first two points will be discussed in this paper primarily. The third point represents an actual physical process and is thus not something numerical adjustments can (or should) influence. As will be shown later in this paper, in some cases, premature magnetic reconnection can actually be triggered by the other aspects, and in such a case, it can be also considered as the result of the artificial electric field rather than its source. The last point refers to the fact that the currently applied mesh, created as a radially-outward extruded icosahedral surface, has locally patches of larger distortion, skewness and non-orthogonality. This can lead to the generation of artificial fluxes, giving rise to mesh-related remnants in the solution. To first separate these mesh-related artefacts, a technique was employed, in which after the first simulation, the mesh was rotated (here by 30 degrees in the positive $\varphi$ direction) and the simulation repeated (see the procedure in \cite{Brchnelova}). This means that in the second simulation, the mesh artefacts occur at different locations in the domain and thus they can be easily identified and corrected for, if needed, \textcolor{black}{using interpolation from the two solutions at the problematic locations.}

\begin{figure*}[t!]
\centering
\gridline{\fig{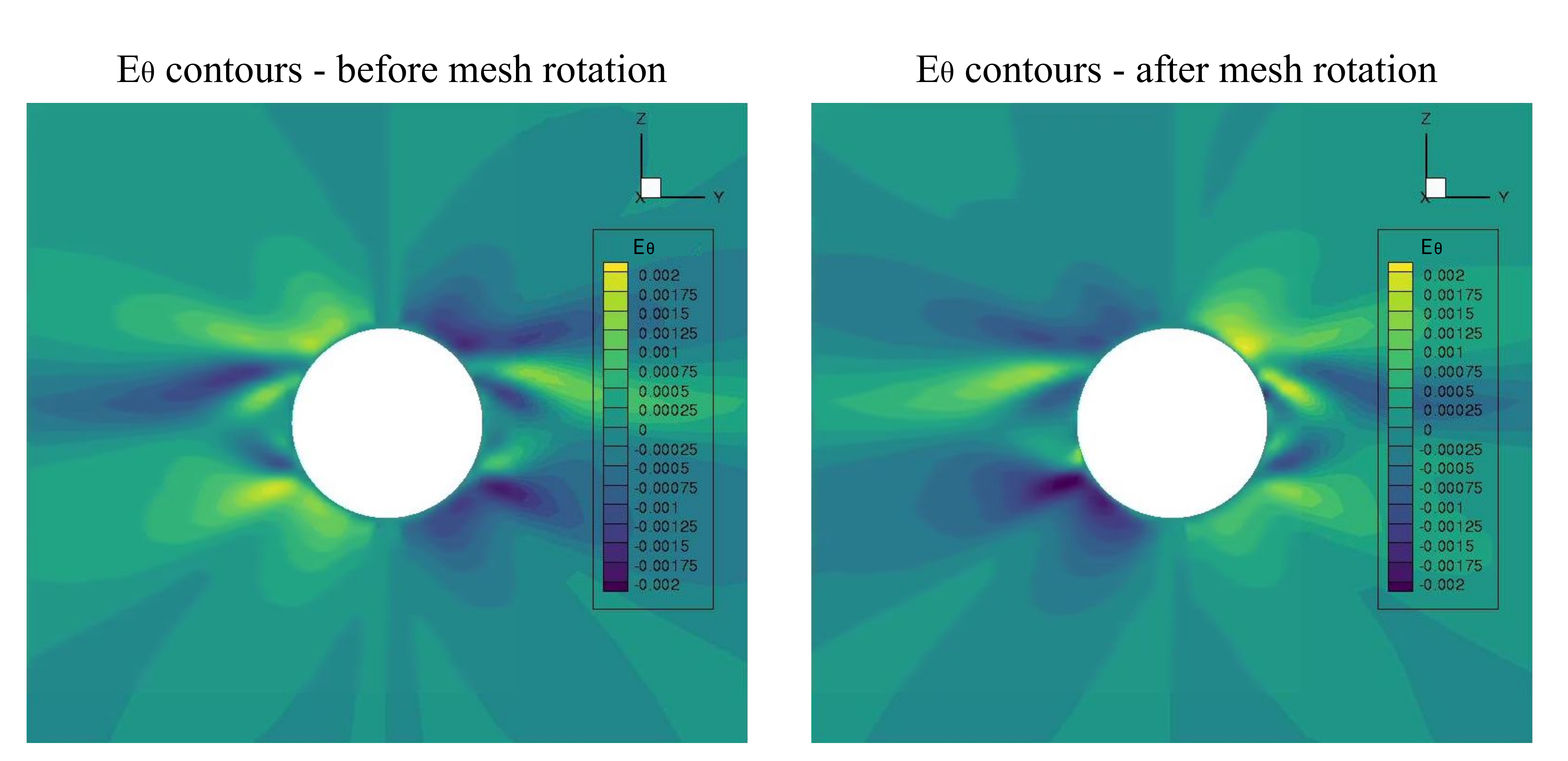}{0.9\textwidth}{}}
\caption{The contours of $E_{\theta}$ before and after mesh rotation, showing that these structures are mostly mesh-related. We show them in a vertical cut (as the dipole does not have longitudinal dependency) with contours set between -0.002 and 0.002.}
\label{fig:Etheta_remnants}
\end{figure*}
In our case, through this technique, it was found that the radial and the azimuthal components of the electric field in the solution of the steady dipole were mostly related to the mesh. An example of the $E_\theta$ remnants is shown in Figure~\ref{fig:Etheta_remnants}, where the electric field changes polarity once the mesh is rotated by 30 degrees in the positive $\varphi$ direction. The same was observed for $E_r$. In general, both of these components had very small values compared to the total $\mathbf{E}$-field magnitude. This is expected when looking at the magnetic field of the non-rotating dipole (which has a negligible $\varphi$-component) and Equation~(\ref{eq:E20}). These two components will thus not be analysed extensively from here on. 

\begin{figure*}[t!]
\centering
\gridline{\fig{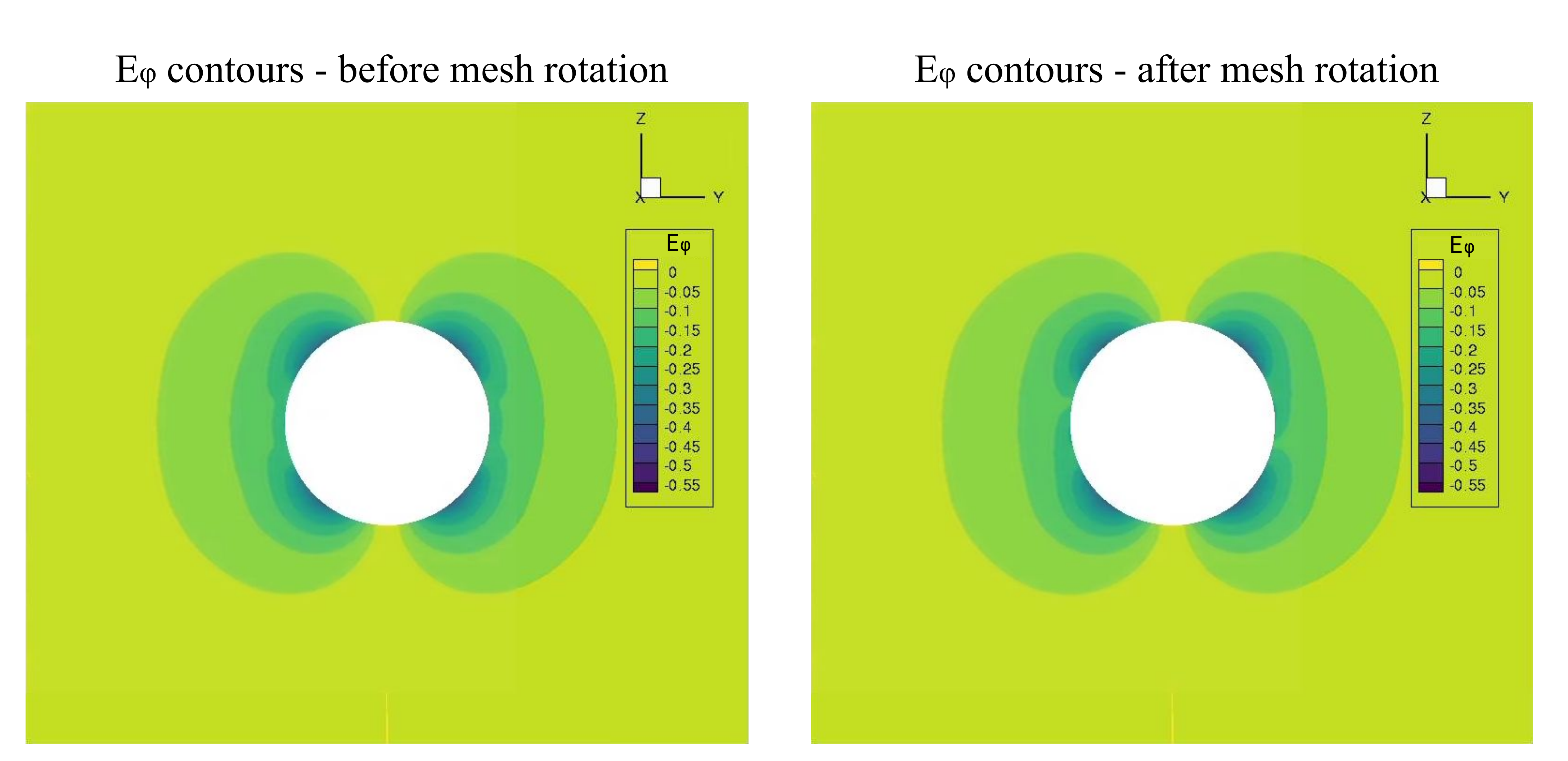}{0.9\textwidth}{}}
\caption{The contours of $E_\varphi$ before and after rotation of the mesh, demonstrating that this $\mathbf{E}$-field component is unrelated to the mesh, but produced through other mechanisms.  We show them in a vertical cut (as the dipole does not have longitudinal dependency) with contours set between -0.55 and 0.0.}
\label{fig:Ephi_remnants}
\end{figure*}

However, the $E_\varphi$ (poloidal) component was a hundred-fold larger and did not change polarity with the rotation of the mesh (see Figure~\ref{fig:Ephi_remnants}), demonstrating that this was created due to reasons other than the mesh. As it is the dominant component, it is the focus of our analysis in the rest of the paper, as far as the dipolar field is concerned. 

\begin{figure*}[t!]
\centering
\gridline{\fig{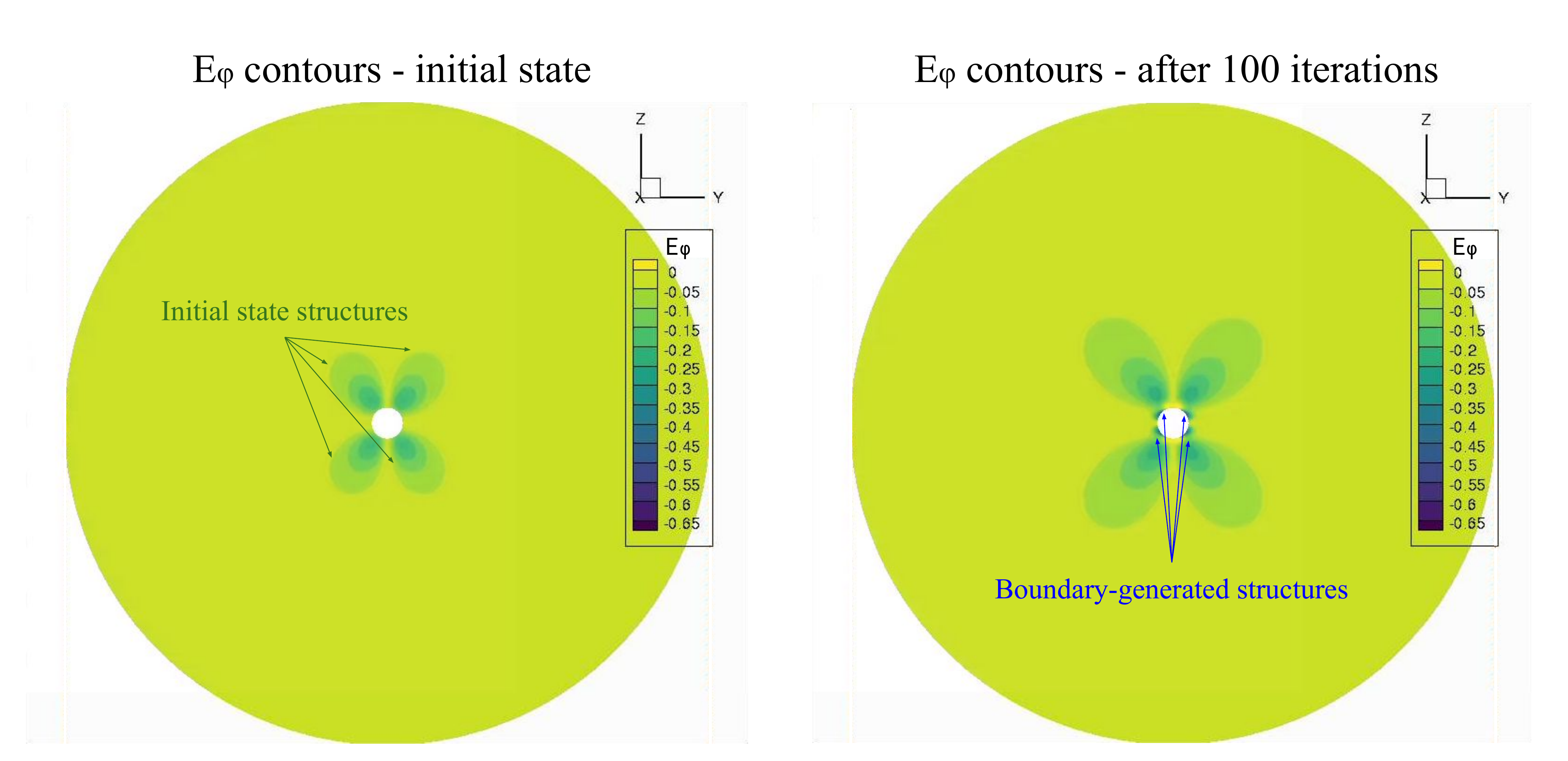}{0.9\textwidth}{}}
\caption{A demonstration of the types of spurious $\mathbf{E}$-field developing within the computational domain: one due to the initial state of the simulation (left, taken at iteration 0) and one caused by the inner BC (right, generated after 100 iterations).}
\label{fig:Ephi_nonstarted}
\end{figure*}

The evolution of $E_\varphi$ during the transient in one of the simulations is shown in Figure~\ref{fig:Ephi_nonstarted}. The features which were introduced at the zero-th iteration (on the left, green arrows), created at the initial state, largely remain in the simulation even after 100 iterations (as shown in the right panel). In this case, the reason why the initial state had a non-zero $\mathbf{E}$-field is the fact that the initial velocity profile came from a piece-wise linear fit of the Parker's wind solution, while the initial $\mathbf{B}$-field came from a Potential Field Source Surface (PFSS) solution, and these two were not parallel. Furthermore, an additional $E_\varphi$ contribution, coming from the BC, develops around the inner boundary during the iterative process towards steady state, as highlighted on the right of Figure~\ref{fig:Ephi_nonstarted} with blue arrows. 
\textcolor{black}{What is shown in Figure~\ref{fig:Ephi_nonstarted} already indicates two possible strategies for reducing the artificial $\mathbf{E}$-field, namely by 1)~ensuring that the initial state is completely $\mathbf{E}$-field-free (set $\mathbf{V}_0$ such that $\mathbf{V}_0 \times \mathbf{B}_\text{PFSS} = 0$), and 2)~adjusting the inner BC to produce minimum contribution to the $\mathbf{E}$-field.}

\subsection{Comparing large-$\mathbf{E}$ and low-$\mathbf{E}$ BC formulations}
\label{sec:effects}

While the fact that the $\mathbf{E}$-field exists and evolves in the simulations has been shown, what it actually means for the steady-state solution has not yet been demonstrated. Whether or not some $\mathbf{E}$-field exists at the initial state depends heavily on the solver, and it is an issue that can be easily fixed in most cases. Therefore, this aspect is not discussed further. Instead, focus will be placed on the formulation of the inner BC.

In this subsection, we will consider two separate cases. The simulation that was shown in the previous paragraphs, with the BC producing the $\mathbf{E}$-field, is the first case, here indicated as the "large $\mathbf{E}$" simulation. The BC here used was taken from the code Wind-Predict \citep{Reville2015, Perri2018} and is one the standard formulations for global coronal models. In this setup, on the inner boundary, it is assumed that the radial and azimuthal components of the $\mathbf{B}$-field can be derived from the PFSS solution, originally expressed in cartesian coordinates:

\begin{equation}
    B_{r,b} =  \frac{{x_{b}}}{r_{b}} {B_{x,\text{PFSS}}} + \frac{{y_{b}}}{r_{b}}  {B_{y,\text{PFSS}}} + \frac{{z_{b}}}{r_{b}}  {B_{z,\text{PFSS}}},
\end{equation}

\begin{equation}
    B_{\theta,b} = \frac{{x_{b} z_{b}}}{\rho_{b} r_{b}} {B_{x,\text{PFSS}}}  + \frac{{y_{b} z_{b}}}{\rho_{b} r_{b}} {B_{y,\text{PFSS}}} - \frac{\rho_{b}}{r_{b}} {B_{z,\text{PFSS}}},
\label{eq:PFSSBtheta}
\end{equation}

\noindent where the $b$ subscript refers to the boundary location and in the instances of coordinate transforms, $\rho_b$ represents $\sqrt{x_b^2 + y_b^2}$. Then, the velocity condition is formulated such that the poloidal flux is removed. Thus, with    

\begin{equation}
    V_{r,b}^* = \frac{1935.07}{(B_{ref}/\sqrt{\mu_0 \rho_{ref}})}, \quad B_{\text{pol}} = \sqrt{B_{r,b}^2 + B_{\theta,b}^2} \quad 
\label{eq:poloidal1a}
\end{equation}

\begin{equation}
\text{and}
\quad V_{||} = V_{r,b}^* \frac{B_{r,b}}{B_{\text{pol}}^2},
\label{eq:poloidal1b}
\end{equation}

\noindent the radial, azimuthal and poloidal velocities on the boundary are:

\begin{equation}
     V_{r,b} = V_{||} B_{r,b}, \quad V_{\theta,b} = V_{||} B_{\theta,b} \quad \text{and} \quad V_{\phi,b} = 0.
\label{eq:poloidal2}
\end{equation}
\noindent Herein, the value of $1935.07$~m.s$^{-1}$ was the prescribed radial outflow on the inner boundary, as coming from the Parker's wind profile. This value was chosen as it is the default value used in the Wind-Predict code, through which our model was originally validated \citep{COCONUT}. 

The other simulation (i.e.\ "low $\mathbf{E}$") was run with the same setup and initial conditions, apart from the fact that on the inner boundary, the $\mathbf{E}$-field was enforced to be zero in all of its components via:

\begin{equation}
    \mathbf{V}_b \times \mathbf{B}_b = \mathbf{0},
\end{equation}

\noindent from which the imposed direction of the velocity on the boundary $\mathbf{V}_b$ was derived. In addition, while in the previous case, the $B_\theta$ was set according to the PFSS solution, see Equation~(\ref{eq:PFSSBtheta}), in this setup, this constraint was omitted, and the $B_\theta$ was determined through a Neumann condition just like $B_\varphi$. 

\begin{figure*}[t!]
\centering
\gridline{\fig{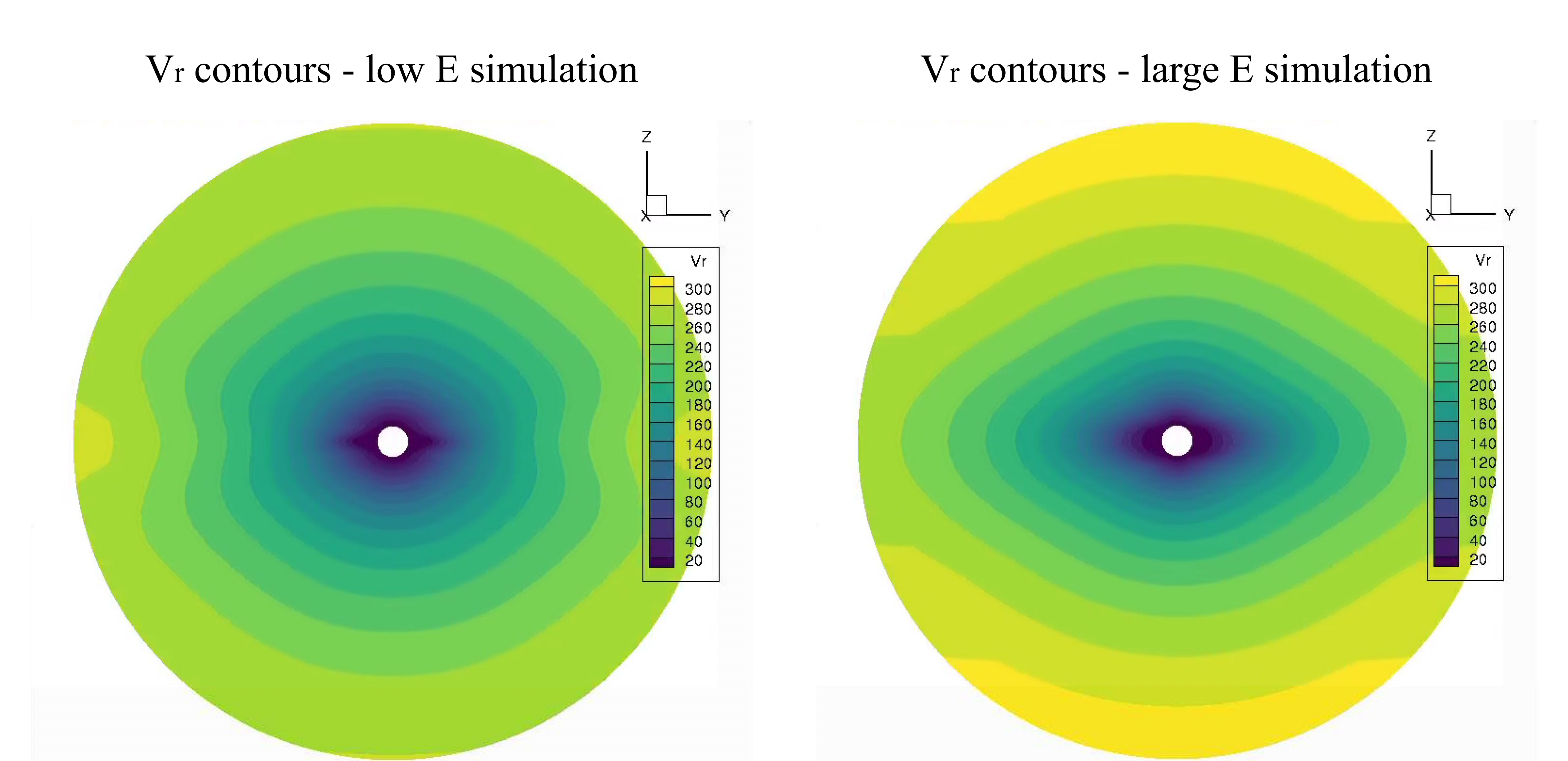}{0.9\textwidth}{}}
\caption{The radial velocity contours of the steady dipole case for two simulation cases with low and large electric fields. For better legibility, units in km/s.}
\label{fig:VrFull}
\end{figure*}

These seemingly small adjustments resulted in considerable differences between the results. The radial velocity contours of the two cases are shown in Figure~\ref{fig:VrFull}. It is apparent that the "large $\mathbf{E}$" case reaches higher velocities on the outer boundary and creates also more diffused profiles of the equatorial streamers. Furthermore, the "low $\mathbf{E}$" simulation has additional features of locally enhanced radial velocity patches near the outer boundary in the equatorial regions. This is a subtle but an important difference, since it fundamentally affects the results on the 21 solar radii surface, which is where the heliospheric EUHFORIA code is supposed to be coupled and take its input data from. \textcolor{black}{These two simulations would have fundamentally different velocity distributions passed to EUHFORIA; one of them with larger velocities near the poles and the other with enhanced velocity in the region of the heliospheric current sheet.}


The observed velocity patches in the equatorial regions are due to magnetic reconnection. \textcolor{black}{The expected profile of the magnetic field lines around a solar dipole with an outflow can be found, for example, in the paper of \cite{Pneuman1971b}. While most of the magnetic field lines remain open, depending on the magnetic field strength, there is a magnetic reconnection forming a current sheet at distances of 1$R_\odot$ to 3$R_\odot$ in the equatorial region. This collimated sheet then propagates throughout the heliosphere. In our simulations however, we see an additional reconnection in this current sheet.}

Due to the finite discretisation and the presence of some unavoidable artificial diffusion, when the current sheet reaches a certain minimum thickness, it reconnects; creating an "X-point" in the $\mathbf{B}$-field lines, opening these lines up behind it. Where exactly (and if) this happens depends on a variety of factors, such as the BCs, the $\mathbf{B}$-field strength and also the grid resolution. The presence of this X-point and its behaviour will be shortly touched upon after the end of this subsection. 

\begin{figure*}[t!]
\centering
\gridline{\fig{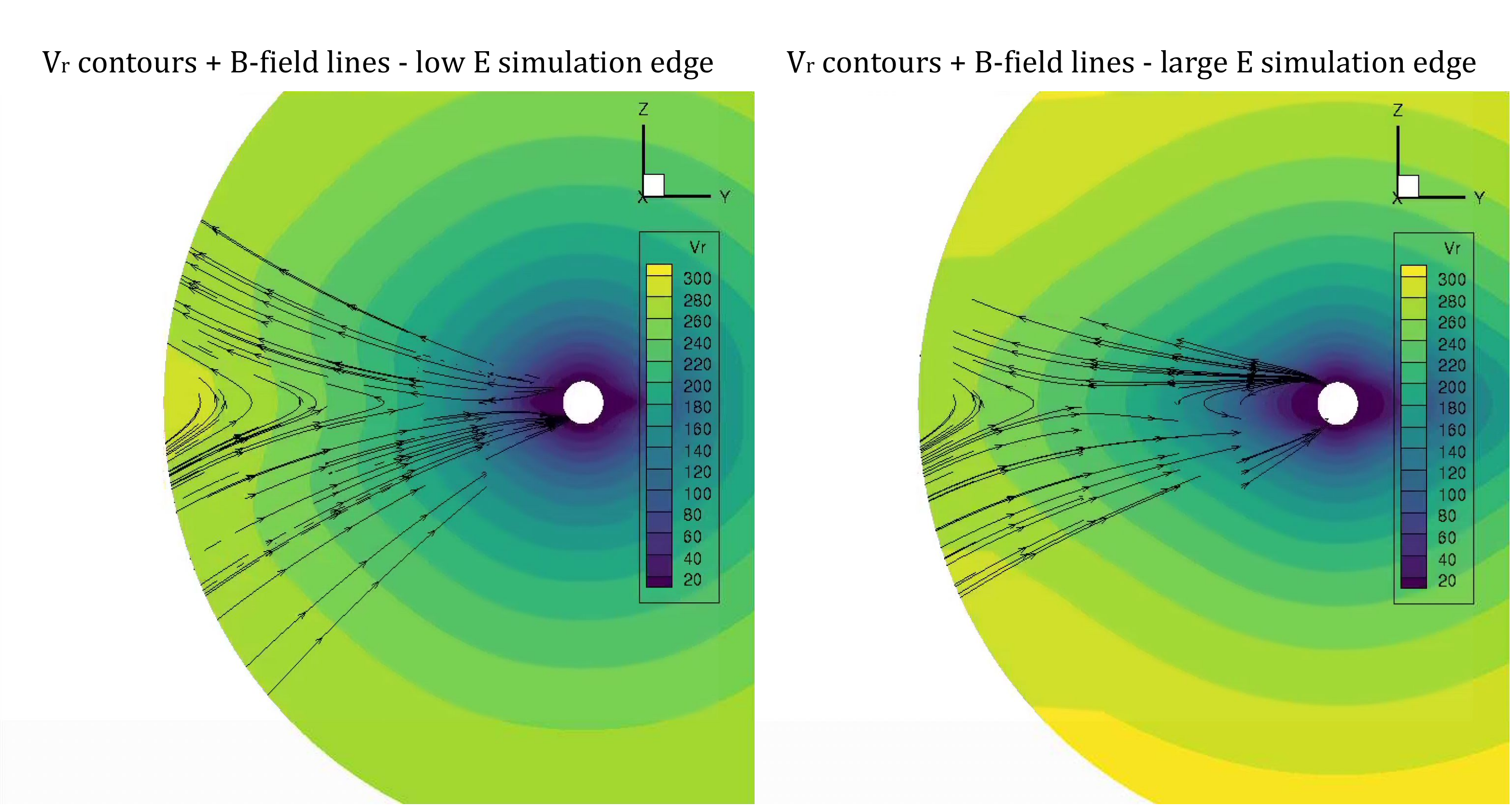}{0.9\textwidth}{}}
\caption{The demonstration of the high velocity patches near the edge of the domain in case of a flow field where the $\mathbf{V}$ and $\mathbf{B}$ distributions are well aligned (on the left) and flow field with a large misalignment (on the right).}
\label{fig:ComparisonsBaloonning}
\end{figure*}

How the flow field responds to this X-point is the cause of the qualitative differences in the velocity profiles in Figure~\ref{fig:VrFull}. In case of the "large $\mathbf{E}$" case, the velocity field does not follow the magnetic field lines well. Thus, even though the $\mathbf{B}$-field changes topology, the velocity profile remains more or less undisturbed as it propagates outwards. As a result, the magnetic reconnection does not show in the final velocity field. For the "low $\mathbf{E}$" run, this is not the case as here, the velocity field does follow the magnetic field, and so it also does respond to the change in its topology. \textcolor{black}{This can be more clearly seen in Figure \ref{fig:ComparisonsBaloonning}, where the edge of the domain is plotted with $V_r$ in the background and with the $\mathbf{B}$-field lines plotted on the top. Behind the X-point, the $\mathbf{B}$-field lines start to open up.} 

\textcolor{black}{This showcased "ballooning" of the streamers due to numerical diffusion is not physical, though it might be stabilising for most simulations. However, it can present problems later on when, for example, determining magnetic connectivity - in case we trace a magnetic field line from the Earth (or another point of interest in the heliosphere) back to the sun, it could get trapped in this diverging region instead of finding its way to the solar surface and to the corresponding photospheric features. In addition, if we project this $B_r$ on the outer surface, in the regions where the current sheet started to diverge, the sign of the $B_r$ field might be incorrect. Additional postprocessing of CFD results might be required to avoid these problems when coupling to heliospheric codes as a result.}

\begin{figure*}[t!]
\centering
\gridline{\fig{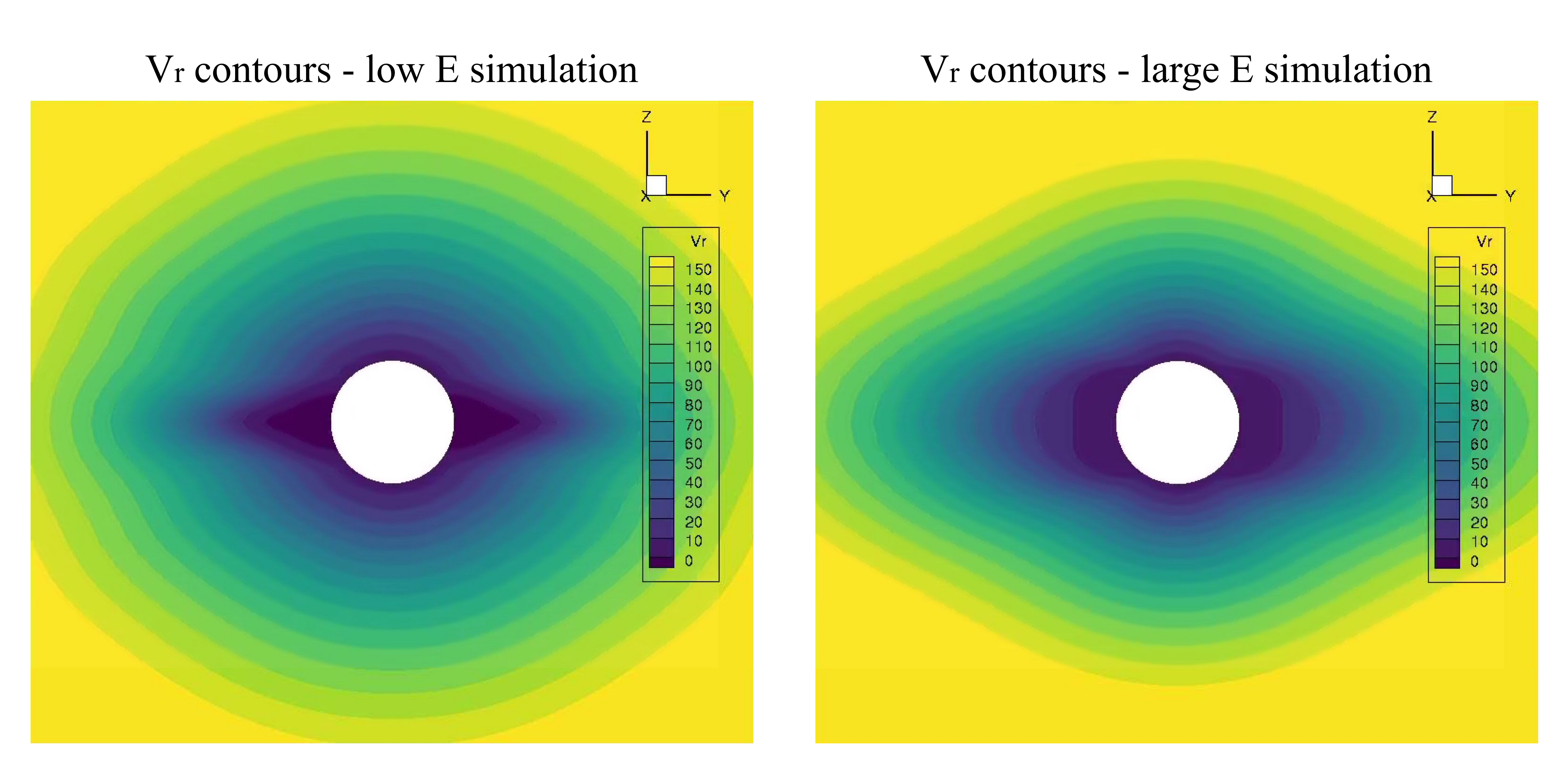}{0.9\textwidth}{}}
\caption{The radial velocity contours near the solar surface, with the colour bar adjusted to demonstrate the enhancement in the sharpness of the streamers for the "low $\mathbf{E}$" simulation (left). For better legibility, units in km/s.}
\label{fig:VrZoom}
\end{figure*}

\begin{figure*}[t!]
\centering
\gridline{\fig{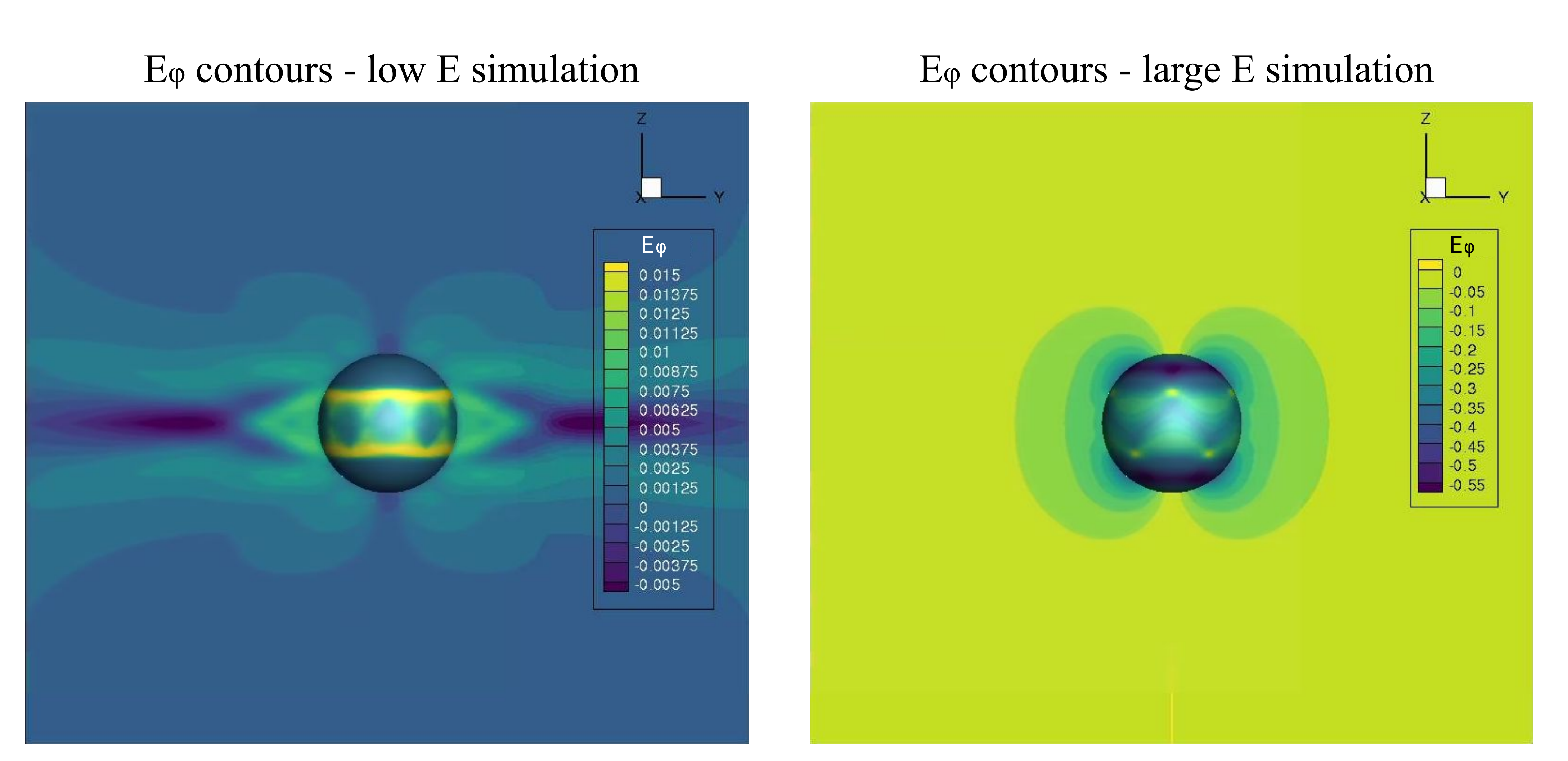}{0.9\textwidth}{}}
\caption{The contours of the $E_\varphi$ field, the colour bar adjusted such that the features shows in both cases.}
\label{fig:Ephi-bothscales}
\end{figure*}


The more diffused profile of the "large $\mathbf{E}$" case, especially around the central region, is shown better as a close-up in Figure~\ref{fig:VrZoom}. Considering that the two simulations use exactly the same grid and none of them uses a flux limiter (which would include excessive numerical dissipation), minimising the $\mathbf{E}$-field leads to a substantial improvement in the resolution of the features. This difference can mean that perhaps a coarser grid could be used in production runs while achieving the same accuracy, which would save computational resources.

In order to visualise how much the $\mathbf{E}$-field has changed between the two cases via the adjustment of the inner BC, in Figure~\ref{fig:Ephi-bothscales}, the $E_\varphi$ is shown. The colourbar of each subplot is adjusted to show the main features.

\begin{figure*}[t!]
\centering
\gridline{\fig{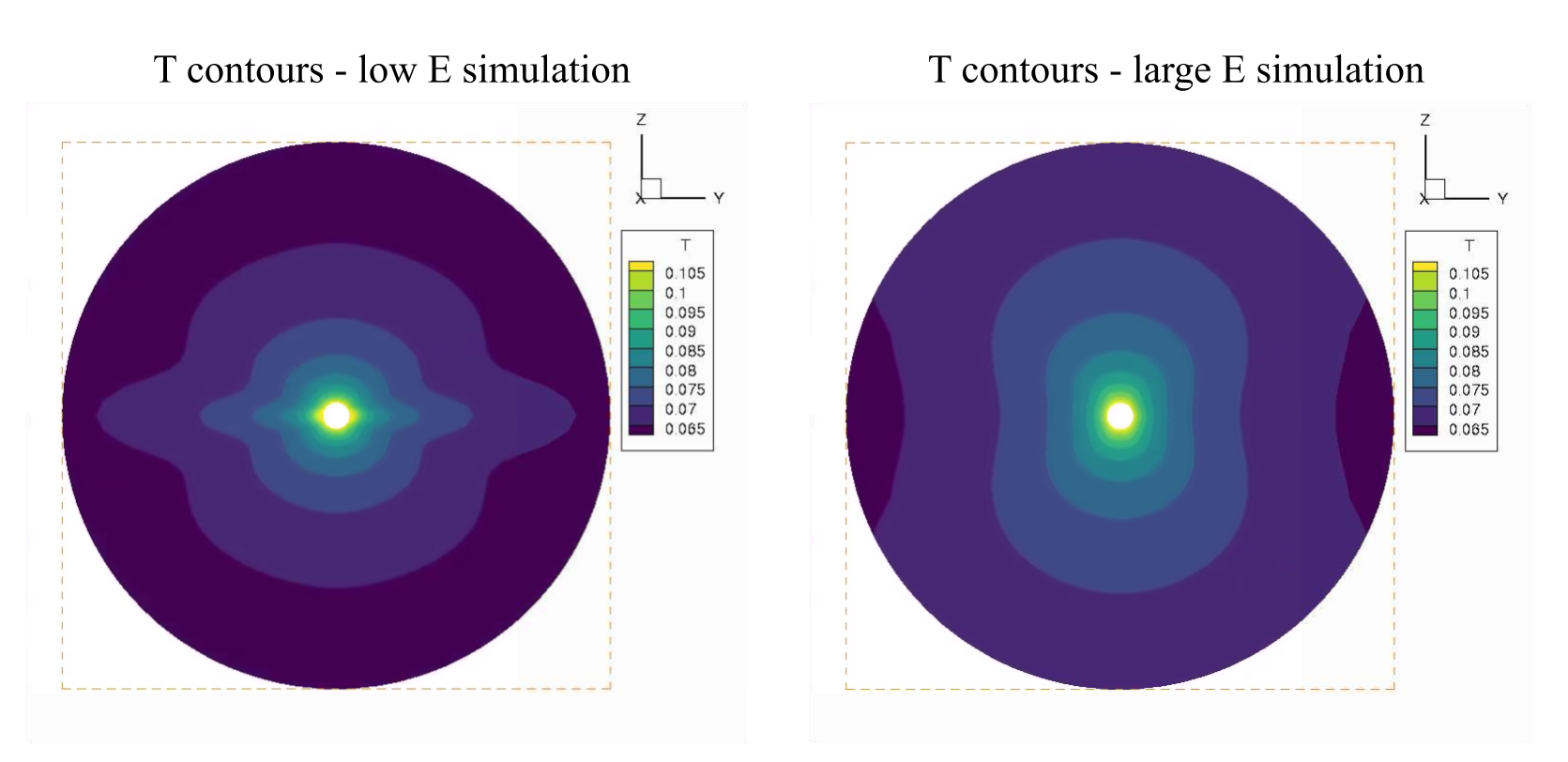}{0.9\textwidth}{}}
\caption{The temperature distribution (here expressed simply as a ratio of pressure and density) of the steady dipole as resolved in the "low $\mathbf{E}$" (left) and "large $\mathbf{E}$" (right) cases.}
\label{fig:TFull}
\end{figure*}

Some regions of a locally larger $\mathbf{E}$-field are still observed in the "low $\mathbf{E}$" case, mainly following the mesh on the surface. Despite that, the $\mathbf{E}$-field is everywhere in the domain at least 50 times smaller than the $\mathbf{E}$-field of the simulation without the adjusted boundary condition (on the right). For the improved case, in the domain, the $\mathbf{E}$-field follows the structure of the streamers, which is where it is the most difficult for the velocity and magnetic fields to align. In the other case, the $\mathbf{E}$-field structure in the domain is mostly spherical, since it is mostly the inner spherical BC generating it.

The sharpness and magnitude of the velocity field are, however, not the only aspects that were found to change significantly through the reduction of the $\mathbf{E}$-field. It was also found that the density and pressure profiles (and thus also the derived temperature profile) were different both qualitatively and quantitatively. Before the adjustment, the temperature (here computed as the ratio of the non-dimensional pressure and density) was enhanced in the regions of strong magnetic field, i.e.\ around the poles for the case of a dipole, see the right side of Figure~\ref{fig:TFull}. After the adjustment, the temperature was instead following the streamers around the equatorial regions (see the left side of the same Figure). 

\begin{figure*}[t!]
\centering
\gridline{\fig{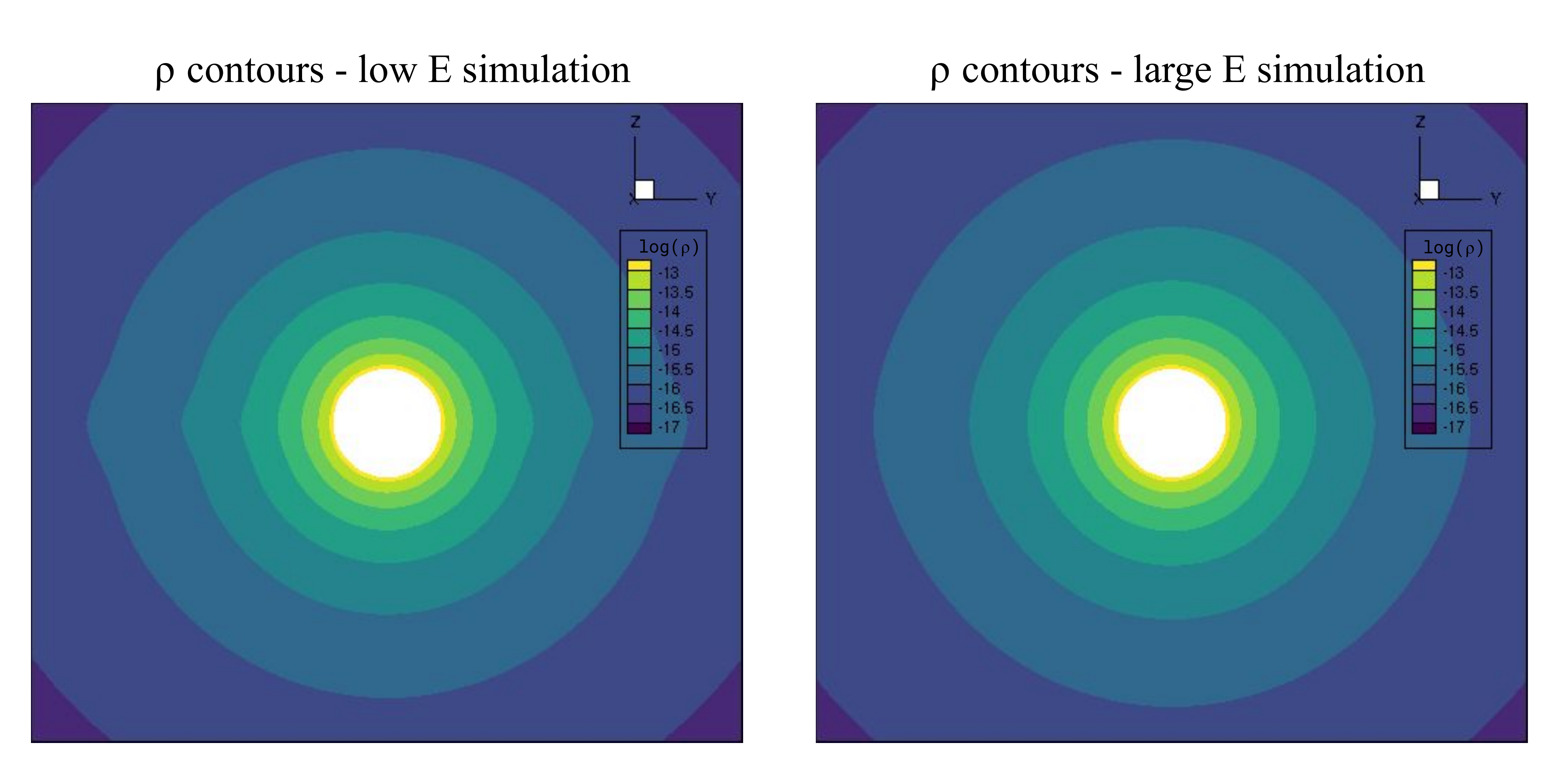}{0.9\textwidth}{}}
\caption{The logarithmic density profiles of the "low $\mathbf{E}$" (left) and "large $\mathbf{E}$" (right) simulations.}
\label{fig:RhoFull}
\end{figure*}

The density profile was also found to be more pronounced for the "low $\mathbf{E}$" simulations, see Figure~\ref{fig:RhoFull}. This might significantly aid the accuracy of procedures such as white-light imaging, where the density is the parameter to be integrated along the line of sight.

\begin{figure*}[t!]
\centering
\gridline{\fig{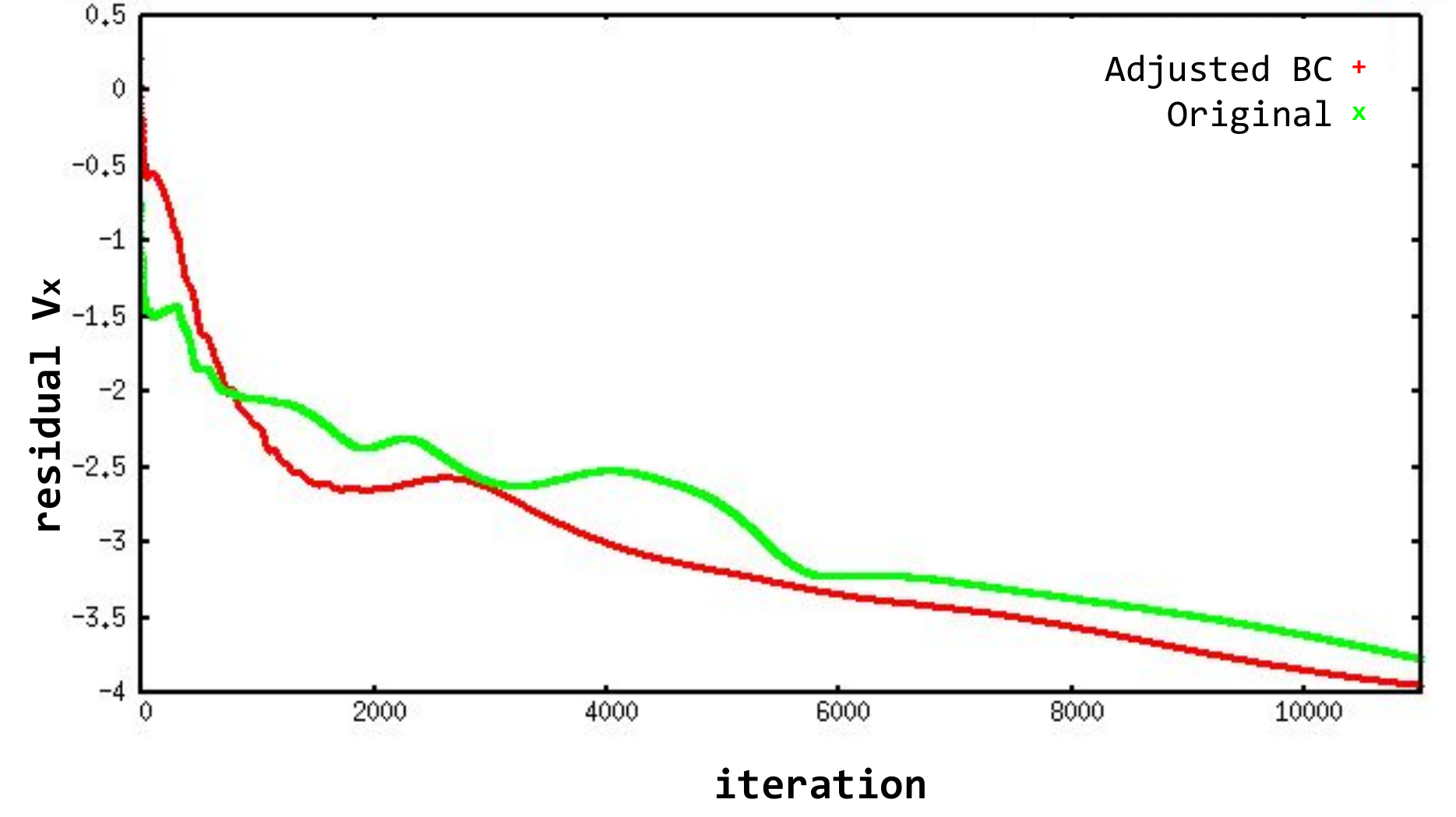}{0.7\textwidth}{}}
\caption{The comparison of the convergence in the $V_x$ component between the two eclipse cases computed with and without an $\mathbf{E}$-field generating boundary.}
\label{fig:ComparisonConvergence}
\end{figure*}

Lastly, convergence impacts should be discussed as computational speed is, next to physical accuracy, a crucial feature of global coronal modelling and any space weather simulation in general. In general, with the new formulation, the steady dipole case required two to three times more iterations in order to reach the same final residual.  As we are mostly interested in realistic magnetic map-driven applications, however, a worse convergence of the dipole case is of relatively little importance. When running a variety of real magnetic map-based cases as discussed in \cite{Kuzma2022}, it was observed that the adjusted BC led to easier convergence when using maps, especially those of the higher $l_\text{max}$. For a constant CFL of 2 for the of the 2008 eclipse used later in this paper, the convergence \textcolor{black}{is} compared for the two inner BC formulations in terms of $V_x$ residual in Figure~\ref{fig:ComparisonConvergence}. After the initial jump in the residual, the two simulations converge to the same residual of -4 after roughly the same number of iterations. The initial larger residuals in the run with the adjusted inner BC is due to the changing $B_\theta$, as it is released from the PFSS solution value.

\subsection{Presence of the X-point and current sheet stability}

Before moving \textcolor{black}{into further investigation of the BCs}, the issue of the presence of the X-point in our CFD simulations, as mentioned in the discussion of Figure~\ref{fig:VrFull}, should be addressed. It is important to discuss this phenomenon also because it provides a context for the analysis of the outer BC formulations, mentioned later on in the paper. 

The (additional) reconnection of the $\mathbf{B}$-field lines which we have observed is normal also in other numerical codes, not just COCONUT (e.g. see the comparison between our code and Wind-Predict in \cite{COCONUT}, where the same behaviour is observed in the solution of Wind-Predict). As found during our numerical experiments, beyond the BCs and the limiter, it is mostly dependent on the grid resolution: with increasing resolution, the reconnection point occurs further away from the Sun. Table~\ref{tab:Xpoint} shows the location of this reconnection point (in terms of solar radii) for four different grids:

\begin{enumerate}
\item The coarse/coarse 300k grid: 5120 surface elements, 65 radial steps
\item The coarse/fine 980k grid: 5120 surface elements, 192 radial steps
\item The fine/coarse 1.3M grid: 20480 surface elements, 65 radial steps
\item The fine/fine 3.9M grid: 20480 surface elements, 192 radial steps
\end{enumerate}

\noindent In Table~\ref{tab:Xpoint}, levels are used in order to express the longitudinal/latitudinal discretisation. Level 5 refers to the 5-th level subdivision of the elementary icosahedron from which the mesh is derived, with 5120 surface elements. Level 6 refers to the 6-th level subdivision with 20480 surface elements. 

By simply refining the domain (without any changes to the physics of the simulation), from Table~\ref{tab:Xpoint}, it can be observed that the X-point moves by almost twice the distance away from the solar surface. Thus, it can be inferred that the reconnection occurs mainly due to the finite discretisation, with its location depending on the amount of numerical diffusion in the domain. 

\begin{table}
\caption{The position of the reconnection point in the domain for a steady dipole (in terms of solar radii) as a function of mesh resolution.}
  \begin{center}
\begin{tabular}{l|l|l|l}
\hline \hline
\#Cells {[}M{]} & Lat./Long. & Radial & X-point {[}solar radii{]} \\ 
0.3             & level 5                  & coarse & 7.2                       \\ 
1.0             & level 5                  & fine   & 8.4                       \\ 
1.3             & level 6                  & coarse & 8.1                       \\ 
3.9             & level 6                  & fine   & 13.3                      \\  \hline \hline
\end{tabular}
  \label{tab:Xpoint}
  \end{center}
\end{table}

\begin{figure*}[t!]
\centering
\gridline{\fig{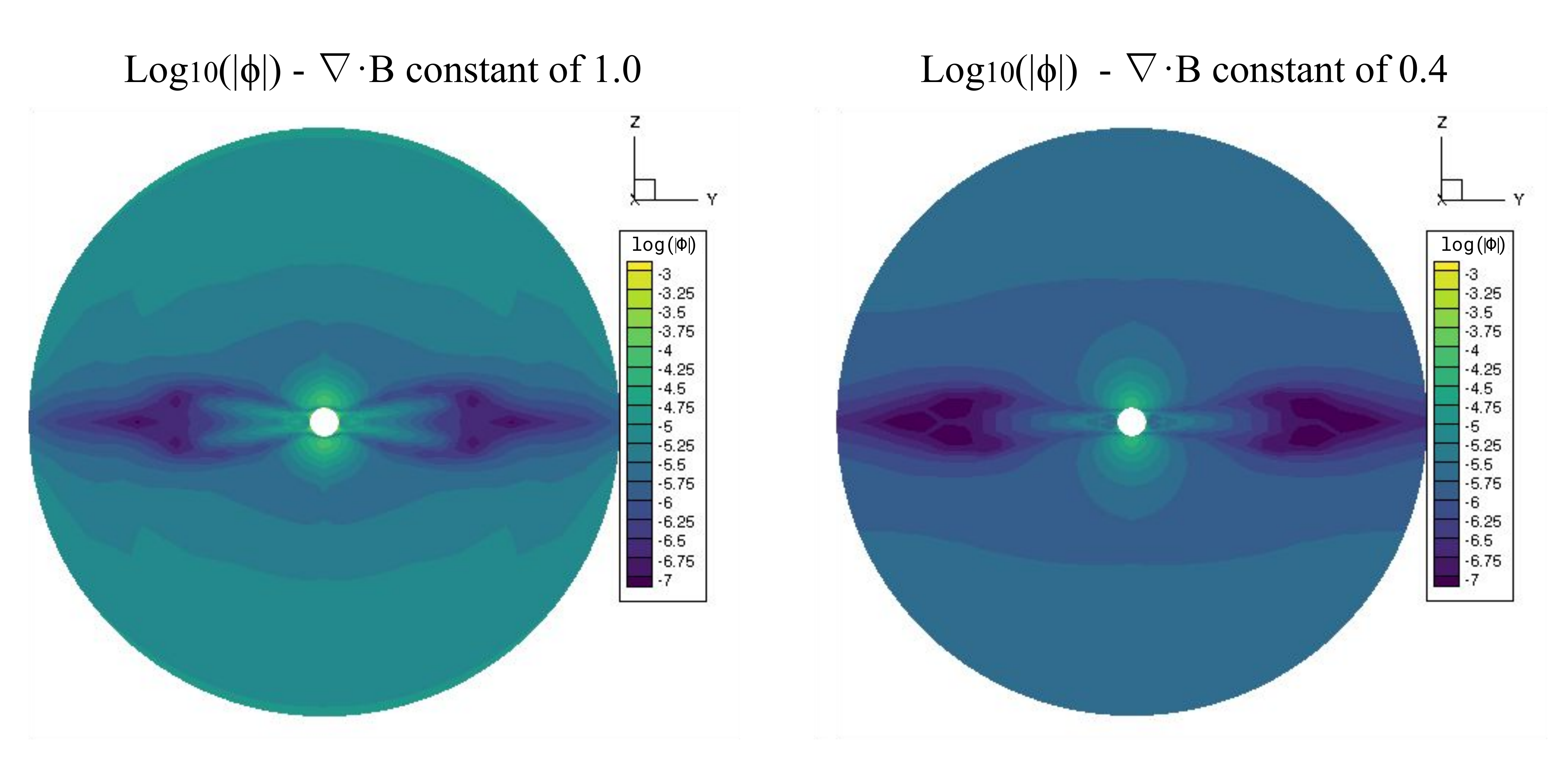}{0.99\textwidth}{}}
\caption{The amount of $\mathbf{\nabla} \cdot \mathbf{B}$ in the numerical solution, depending on the selection of the divergence cleaning constant (1 on the left, 0.4 on the right), is shown in terms of the logarithm of the absolute value of $\mathbf{\nabla} \cdot \mathbf{B}$ (non-dimensional).}
\label{fig:CurrentSheetLogdivB}
\end{figure*}

In addition, it was found that the behaviour and shape of the current sheet (and thus also the location of the reconnection point) also depends on the amount of the $\mathbf{B}$-field divergence in the flow field. By default, in our simulations, for the $\mathbf{B}$-field divergence cleaning constant, we use the value of 1, leading to the amount of divergence of $\mathbf{B}$ to be in the order of -4 to -5 (nondimensional). With lower values of the cleaning constant, it was observed that the amount of the divergence of $\mathbf{B}$-field decreased, see Figure~\ref{fig:CurrentSheetLogdivB}, where the logarithm of the absolute value of $\phi$ (the divergence cleaning parameter) is shown. The BC treatment for $\phi$ (homogeneous Neumann or Dirichlet) on the inner boundary does not significantly affect this distribution and in the default configuration, $\phi$ is set to 0 (i.e. homogeneous Dirichlet). 

\begin{figure*}[t!]
\centering
\gridline{\fig{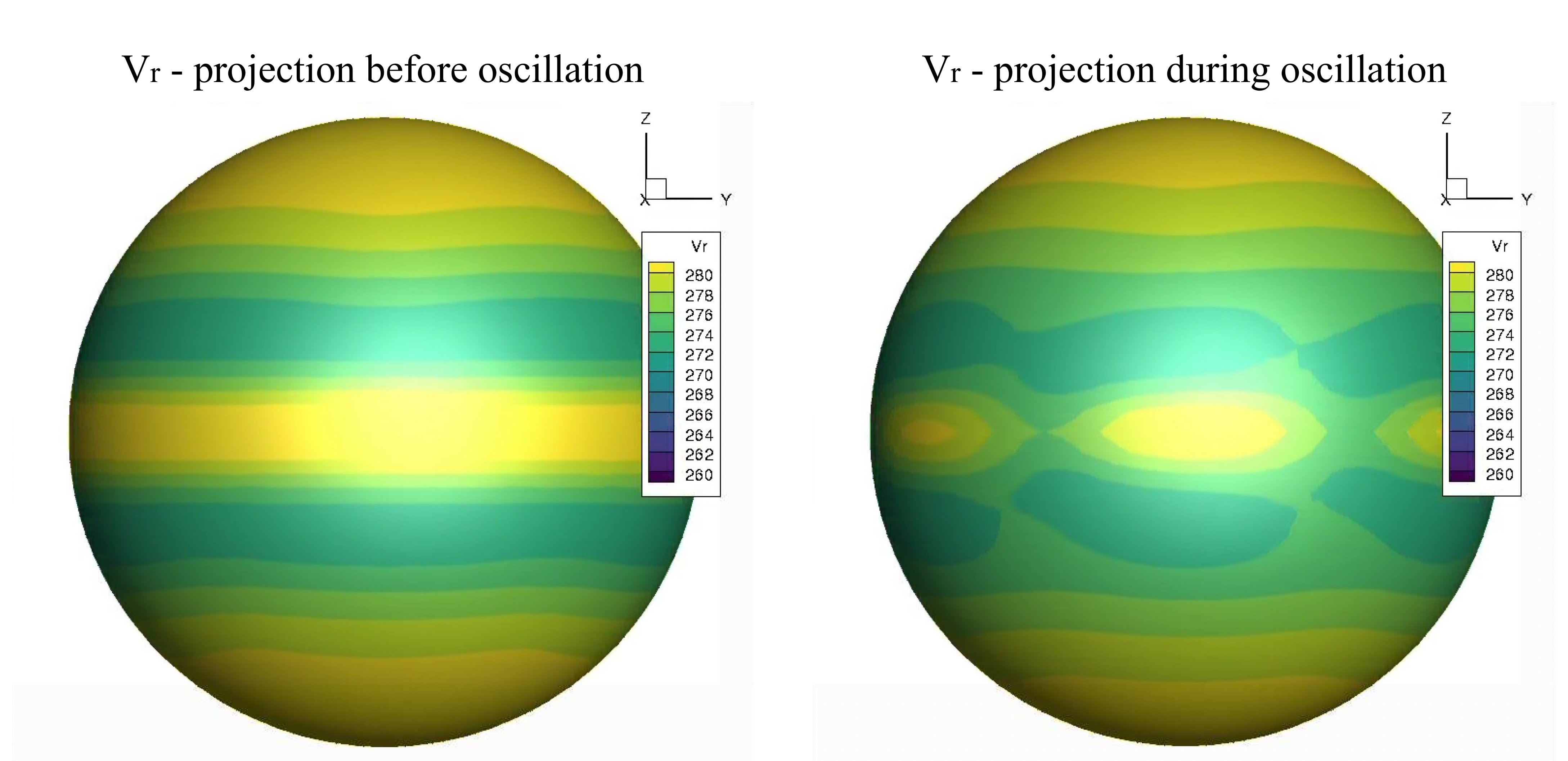}{0.9\textwidth}{}}
\caption{The projection of $V_r$ in $[m.s^{-1}]$ on the outer surface of the domain before (left) and during (right) the unstable-current-sheet residual oscillation shows that once the unstable oscillation is activated, the current sheet stops being continuous and forms mass lumps that are ejected outwards.}
\label{fig:CurrentSheet}
\end{figure*}

However, decreasing the value of the cleaning constant below a certain threshold destabilises the simulation. The particular value which causes this most likely depends on the specific scheme, the mesh and the BCs; in our case it is around the value of 0.3. The simulation reaches a certain minimum residual, following which the residual increases again and starts to oscillate indefinitely. What this means for the simulation is that the current sheet becomes unstable. The reconnection region starts to move forward and backward between iterations (on the current mesh, generally oscillating between 5 and 10 solar radii), creating lumps of mass which are ejected outwards. To illustrate this, the projection of the radial velocity on the outer surface (where the passing lumps are the most visible) is shown in Figure~\ref{fig:CurrentSheet}. We should note that this kind of unstable structures is actually visible in the observations of the solar wind, although they appear for more physical reasons (reconnection along the HCS that eject flux ropes and density structures, see \cite{Reville2020}).

Thus, it it apparent at this point that both the numerical dissipation due to the finite discretisation and the amount of $\mathbf{\nabla} \cdot \mathbf{B}$ in the domain affect the shape and behaviour of the current sheet. As we aim at developing software which will eventually have standardised and automatised handling of magnetic map-based simulations, the above-mentioned oscillating behaviour in the residual is not desirable, even though at certain phases, the reconnection point is located much further away from the Sun than without it. When the residual starts oscillating, it becomes difficult to infer the state of  convergence of the solution without any visual inspection. Since the higher values of the cleaning constant give quantitative results comparable to the time-averaged values of the oscillating solution, it was decided to keep the value of the cleaning constant to 1 for all our simulations. 

Even though we can now explain the behaviour and the origin of the X-point, its presence still complicates matters, for reasons already outlined earlier such as coupling to EUHFORIA or magnetic connectivity studies. 
It is thus important to lower the amount of numerical diffusion as much as computationally feasible and to move this reconnection point as close to the outer boundary as we can; which can be, to some extent, realised through the outer boundary conditions. 

\section{Formulating the Outer BC through a Parametric Study}
\label{sec:outer}
It has already been established that the inner BC has a major influence on the results. In order to maximise our gain in terms of accuracy and computational performance for a given mesh resolution, the outer BC should also be addressed. 

Firstly, a parametric study will be presented, where the formulations of all of the primitive variables are varied and studied systematically. Afterwards, special focus will be placed on how the density and magnetic field BCs should be formulated at 21.5$R_\odot$. Finally, a new technique will be demonstrated where the BC effects can be further reduced via a careful grid design. 

\subsection{Parametric outer BC analysis}

As pointed out in the introduction, the formulation of the outer BC in available literature is often reduced to calling it a simple extrapolation. It is indeed clear that since the outer boundary is supersonic, the hydrodynamic variables cannot be prescribed, but they must be extrapolated from the inner states. However, this extrapolation can be formulated in various ways. The prescription can simply be done through a zero-gradient extrapolation (i.e. homogeneous Neumann condition) or through the extrapolation of the given variable via a function of the radius or density. The effects that these different formulations have on the behaviour of the code or shape of the solution has, to the Authors' knowledge, not yet been extensively discussed in the available literature. For this reason, a parametric study was performed to analyse these effects in a systematic way. 

\begin{table*}[]
\centering
\caption{The setup of 9 different dipole simulation cases with varying outer BC settings. }
\begin{tabular}{c|c|c|c|c|c|c|c|c|c}
No & $\rho$                                           & $B_r$                                                & $B_\theta$ & $B_\varphi$        & $V_r$                                                             & $V_\theta$           & $V_\varphi$        & $p$                                                 & $\phi$   \\ \hline
1  & G = I                                             & G = I                                             & G = I  & G = I       & G = I                                                          & G = I            & G = I       & B = I                                             & B = 0 \\  \hline
2  & G = I $\frac{r^2_I}{r^2_G}$ & G = I                                             & G = I  & G = I       & G = I                                                          & G = I            & G = I       & G = I $\frac{r^2_I}{r^2_G}$ & B = 0 \\ \hline
3  & G = I $\frac{r^2_I}{r^2_G}$ & G = I $\frac{r^2_I}{r^2_G}$ & G = I  & G = I       & G = I                                                          & G = I            & G = I       & G = I $\frac{r^2_I}{r^2_G}$ & B = 0 \\  \hline
4  & G = I $\frac{r^2_I}{r^2_G}$ & G = I $\frac{r^2_I}{r^2_G}$ & G = I  & G = I $\frac{r_I}{r_G}$ & G = I                                                          & G = I            & G = I       & G = I $\frac{r^2_I}{r^2_G}$ & B = 0 \\ \hline
5  & G = I $\frac{r^2_I}{r^2_G}$ & G = I $\frac{r^2_I}{r^2_G}$ & G = I  & G = I $\frac{r_I}{r_G}$ & G = I                                                          & G = I            & G = I       & G = I $\frac{r^2_I}{r^2_G}$ & G = I \\  \hline
6  & G = I $\frac{r^2_I}{r^2_G}$ & G = I $\frac{r^2_I}{r^2_G}$ & G = I  & G = I $\frac{r_I}{r_G}$ &G = I $\frac{r^2_I \rho_I}{r^2_G \rho_G}$ & G = I  & G = I $\frac{r_I}{r_G}$ & G = I $\frac{r^2_I}{r^2_G}$ & B = 0 \\ \hline
7  & G = I $\frac{r^2_I}{r^2_G}$ & G = I $\frac{r^2_I}{r^2_G}$ & G = I  & G = I       &G = I $\frac{r^2_I \rho_I}{r^2_G \rho_G}$ & G = I & G = I $\frac{r_I}{r_G}$ & G = I $\frac{r^2_I}{r^2_G}$ & B = 0 \\  \hline
8  & G = I $\frac{r^2_I}{r^2_G}$ & G = I                                             & G = I  & G = I       &G = I $\frac{r^2_I \rho_I}{r^2_G \rho_G}$ & G = I & G = I $\frac{r_I}{r_G}$ & G = I $\frac{r^2_I}{r^2_G}$ & B = 0 \\  \hline
9  &  G = I $\frac{r^3_I}{r^3_G}$ & G = I $\frac{r^2_I}{r^2_G}$ & G = I  & G = I $\frac{r_I}{r_G}$ &G = I $\frac{r^2_I \rho_I}{r^2_G \rho_G}$ & G = I  & G = I $\frac{r_I}{r_G}$ &  G = I $\frac{r^3_I}{r^3_G}$ & B = 0 \\ \hline \hline
\end{tabular}

\label{tab:casesouter}
\end{table*}

All of the primitive variables of the code, $\rho, B_x, B_y, B_z, V_x, V_y, V_z, p, $ and $\phi$ (the divergence cleaning parameter), were extended either via zero-gradient (the ghost state being equal to the inner state, where the distances of the inner cell centre and the ghost cell centre to the boundary are equal) or via a function of radius or density, or both. The formulations of the specific functions for the velocity and the magnetic field were taken (apart from the last expression in Equation~(\ref{eq:rho})) from \cite{Talpeanu}. The following expressions for density

\begin{equation}
    \rho = \text{cst.} \quad \text{or} \quad \rho r^2 = \text{cst.}   \quad \text{or} \quad \rho r^3 = \text{cst.},
\label{eq:rho}
\end{equation}

\noindent velocity

\begin{equation}
    V_r =  \text{cst.} \quad V_\theta =  \text{cst.} \quad V_\varphi =  \text{cst.} \quad 
\end{equation}
or
\begin{equation}
    V_r r^2 \rho =  \text{cst.} \quad V_\theta =  \text{cst.} \quad V_\varphi r =  \text{cst.},
\end{equation}

\noindent magnetic field
\begin{equation}
    B_r =  \text{cst.} \quad B_\theta =  \text{cst.} \quad B_\varphi =  \text{cst.} \quad 
\end{equation}
or
\begin{equation}
    B_r r^2 =  \text{cst} \quad B_\theta =  \text{cst.} \quad B_\varphi r =  \text{cst.}
\end{equation}

\noindent and $\phi$

\begin{equation}
    \phi = \text{cst.} \quad \text{or} \quad \phi_b = 0,
\end{equation}

\noindent were combined into 9 different cases, as summarised in Table~\ref{tab:casesouter}. In the Table, "G" refers to the ghost state, "I" to the inner state and "B" to the location on the boundary. To achieve continuity in temperature, the pressure was extrapolated in the same way as the density. The magnetic configuration was again dipolar for easier evaluation. All of these cases were simulated with the same CFL law (starting from 2 with the CFL doubling every few hundred iterations up to 128) for controlling the convergence to steady state, making comparisons between them possible. 

\begin{table}
\centering
\caption{Results for the X-point location, the maximum $V_r$ and the time to converge down to the residual of -10 in $V_x$ for the various outer BC simulation cases.}
\begin{tabular}{l|l|l|l}
\hline \hline
No & X-point [$R_\odot$] & $V_{r, \text{max}}$ [ms$^{-1}$] & $T_{V_x \text{res} < -10}$ [s]  \\ \hline
1  & 5.8    & 3.09E+05 & 7135.12 \\
2  & 6.7    & 2.87E+05 & 7174.89 \\
3  & 7.2    & 2.86E+05 & 7150.49 \\ 
4  & 7.2    & 2.84E+05 & 7162.78 \\ 
5  & 6.5    & 2.84E+05 & N/A  \\ 
6  & 7.2    & 2.84E+05 & 7171.25\\ 
7  & 7.2    & 2.84E+05 & 7157.08 \\
8  & 7.1    & 2.86E+05 & 7161.79\\
9  & 7.2    & 2.79E+05 & 7157.28\\ \hline \hline
\end{tabular}
\label{tab:outerresults}
\end{table}

For each simulation, the $V_r$ profiles were compared as well as the maximum $V_r$ value and the location of the X-point in the $\mathbf{B}$-field lines, \textcolor{black}{since its location generally directly corresponded to the divergence of the field lines near the outer boundary}. The evaluation of the exact location of the X-point has uncertainty of approximately $\pm 0.05 R_\odot$, since it was determined by eye inspection of the solution in a visualisation software. In addition, the total time of each simulation required to reach the $V_x$ residual of -10 was also measured to compare the convergence. The results for the 9 cases are given in Table~\ref{tab:outerresults}. 

From observing the convergence alone, it can be quickly concluded that the divergence cleaning parameter $\phi$ must be set to 0 via the Dirichlet boundary condition instead of being extrapolated via zero-gradient, given that the fifth simulation never converged. The results presented in Table~\ref{tab:outerresults} were taken at the $V_x$ residual of roughly -2.5. It was further experimented to see which Dirichlet formulation is better - setting $\phi_g = 0$ or $\phi_b = 0$ - but both led to roughly the same convergence and the same solution, with the former giving slightly larger divergence of the $\mathbf{B}$-field lines near the outer boundary. The convergence of the other cases, in terms of the computational resources needed, was almost the same.  

According to the results in Table~\ref{tab:outerresults}, if we consider the other two evaluated parameters alone (the maximum $V_r$ and the location of the X-point), the following points can be further raised:

\begin{itemize}
    \item from comparing cases 3 \& 4 and 6 \& 7: whether we extrapolate $B_\varphi$ via zero-gradient or radius is of no consequence (at least for the non-rotating dipole, where $B_\varphi$ is negligible);
    \item from comparing cases 7 \& 8 and 8 \& 9: $B_r$ should be extended via a function of radius (here $r^{-2}$) instead of a zero-gradient (this will be revisited in the subsection below);
    \item from comparing cases 2 \& 8 and 4 \& 6: whether we extrapolate $\mathbf{V}$ via zero-gradient or the given laws is of barely any consequence;
    \item from comparing cases 1 \& 2: the density (and pressure) should be extrapolated as a function of radius (here $r^{-2}$);
    \item from comparing cases 6 \& 9: the density (and pressure) extended via $r^{-3}$ gives lower maximum speeds.
\end{itemize}


\begin{figure*}[t!]
\centering
\gridline{\fig{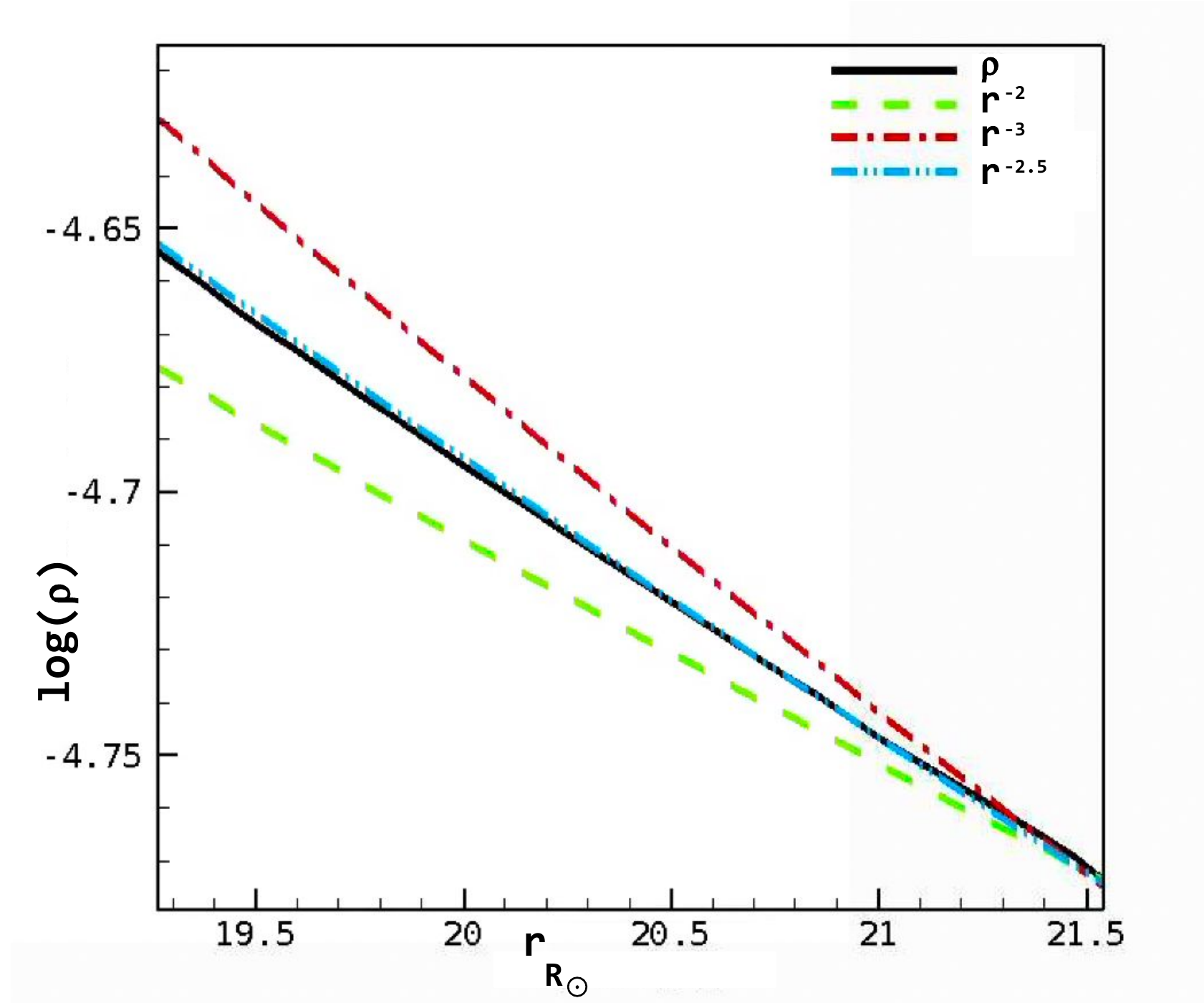}{0.6\textwidth}{}}
\caption{The comparison of logarithms of different distance functions (inverse cubic in red, inverse square in green and inverse 2.5 power-law profile in blue) with the behaviour of the logarithm of the nondimensional density versus the distance from the Sun (in solar radii). The profiles were designed to intersect at 21.5$R_\odot$ to allow for slope analysis.}
\label{fig:logrhov4}
\end{figure*}

Obviously, simply observing the location of the X-point alone is insufficient to justify the variable prescription to make the case physical. Nevertheless, these results are a good indication to which conditions the behaviour of the solution is very sensitive and to which it is not. Especially the extrapolation of the radial magnetic field component seem to have a profound effect on the magnetic field line topology, judging from the shifting of the reconnection location. In addition, the formulation of the extrapolation of density also affects the maximum speeds in the domain. Thus, these two aspects will be elaborated on in more detail in the following subsections.

\subsection{Density profile analysis}

While  Table~\ref{tab:outerresults} indicates that the density extrapolation affects the maximum speed in the domain it is still not clear whether a more physically accurate result is given by the inverse square, inverse cube or perhaps some other law. The most of the available coronal models use the inverse square extension (see, for example, \cite{Talpeanu}), \textcolor{black}{ as this represents a flow that is expanding spherically with a uniform radial speed. However, from the $V_r$ contours shown in, for example, Figure \ref{fig:VrFull}, it is apparent that the flow is still accelerating near the outer boundary, making the uniform-flow assumption incorrect. In addition, it will be shown later that in some regions of the domain of data-driven cases, the flow is expanding super-radially. For these reasons, a steeper descent of $\rho$ could be expected.}

In order to determine which formulation is more accurate, an extended-grid analysis was performed, where the computational domain radius was set to 30$\;R_\odot$ and the cells around 21.5 solar radii were more refined radially to capture the local gradients better. This way, the outer BC was relatively far away from this region, and thus the undisturbed density profile at 21.5$\;R_\odot$ that would exist at this location without the presence of the boundary could be investigated. The setup was otherwise equivalent to the case 9 from the parametric study described above.

Initially, $1/r^2$ and $1/r^3$ profiles were compared to the non-dimensional density. For easier comparison of the local slopes, the logarithms of the profiles were used for the analysis. These logarithms were plotted around the region of interest, at 21.5$R_\odot$ where they were made to intersect with the logarithmic non-dimensional density profile as extracted from the CFD solution. The results are shown in Figure~\ref{fig:logrhov4}, where the green profile is the inverse square distance function and the red profile the inverse cubic distance function.   

It was observed that the actual CFD density slope lied roughly in the middle between, and thus also a new profile, the function of $1/r^{-2.5}$ was added, shown as the blue line in Figure~\ref{fig:logrhov4}. It can be concluded easily that this profile matches the density slope at the given radius almost perfectly. With this power law used as the outer BC for density, the maximum reached radial velocity for the steady dipole (now again modelled in the 21.5$\;R_\odot$ domain) is 2.82E+05m.s$^{-1}$, considering the same resolution as used for the cases in Table~\ref{tab:casesouter}. 

\subsection{Magnetic field analysis}

\begin{figure*}[t!]
\centering
\gridline{\fig{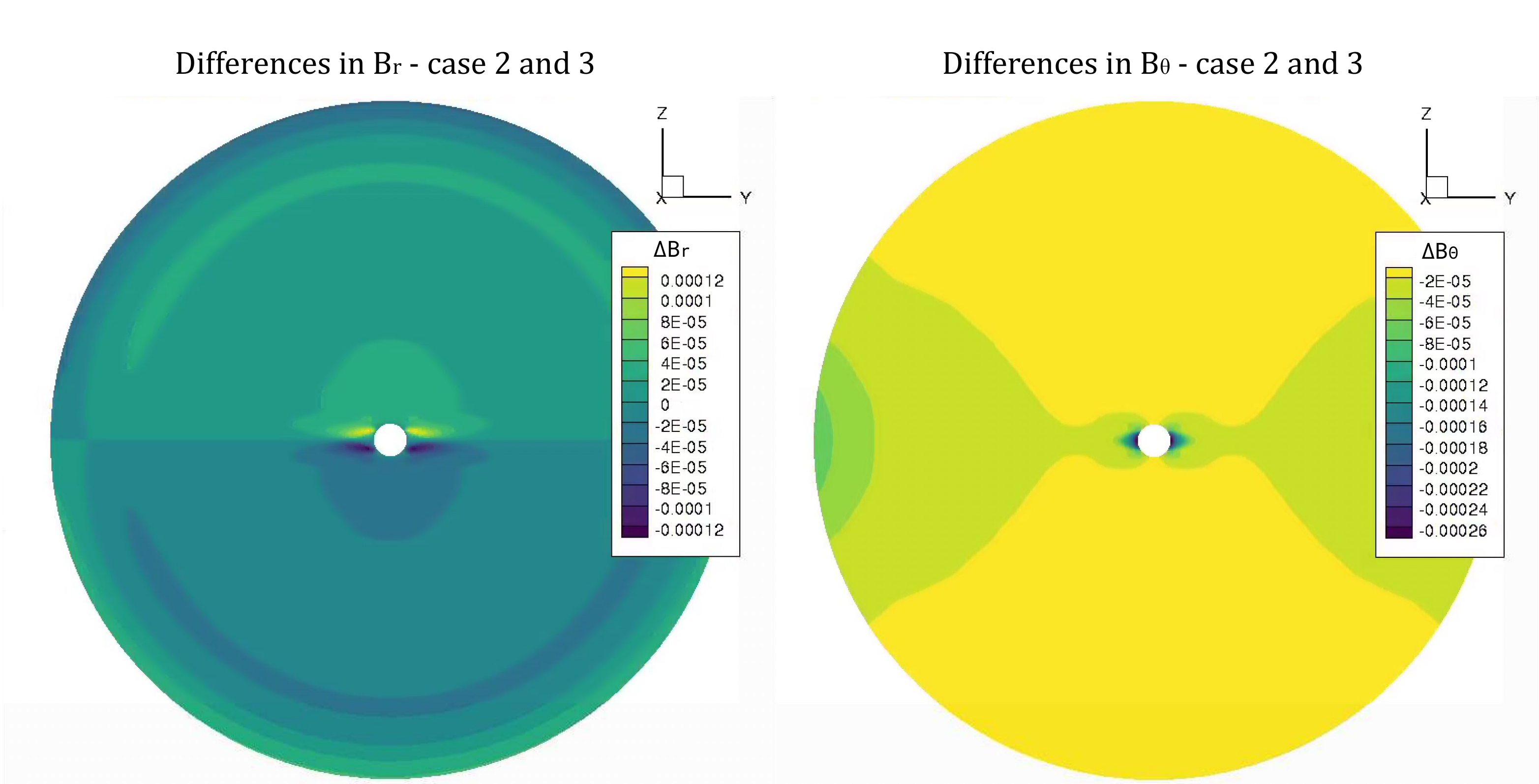}{0.9\textwidth}{}}
\caption{The differences  $B_{r, \text{diff}} = B_{r,2} - B_{r,3 }$ (left) and $B_{\theta, \text{diff}} = B_{\theta,2} - B_{\theta,3 }$ (right) between cases 2 and 3 as a result of using different $B_r$ extrapolations.}
\label{fig:ComparisonBrBtheta}
\end{figure*}

\textcolor{black}{
The same analysis was performed on the magnetic field BC. For a flow that is not expanding super-radially, the magnitude of $B_r$ should be roughly decreasing with $r^2$. Capturing the correct gradient of $B_r$ is important as it also strongly affects how the $B_\theta$ and $B_\phi$ are also resolved. To demonstrate the difference that the extrapolation of $B_r$ alone causes, we compare $B_r$ and $B_\theta$ between the cases 2 and 3 from Table \ref{tab:casesouter} by evaluating the differences $B_{r, \text{diff}} = B_{r,2} - B_{r,3 }$ and $B_{\theta, \text{diff}} = B_{\theta,2} - B_{\theta,3 }$ in Figure \ref{fig:ComparisonBrBtheta}. From this Figure, two observations can be made. Firstly, as a result of the outer BC formulation, $B_r$ changes not only near the boundary, but also inside of the domain. Secondly, it is apparent that also $B_\theta$ changes as a result of the $B_r$ extrapolation, despite the fact that the condition for $B_\theta$ remained unchanged. }

\begin{figure*}[t!]
\centering
\gridline{\fig{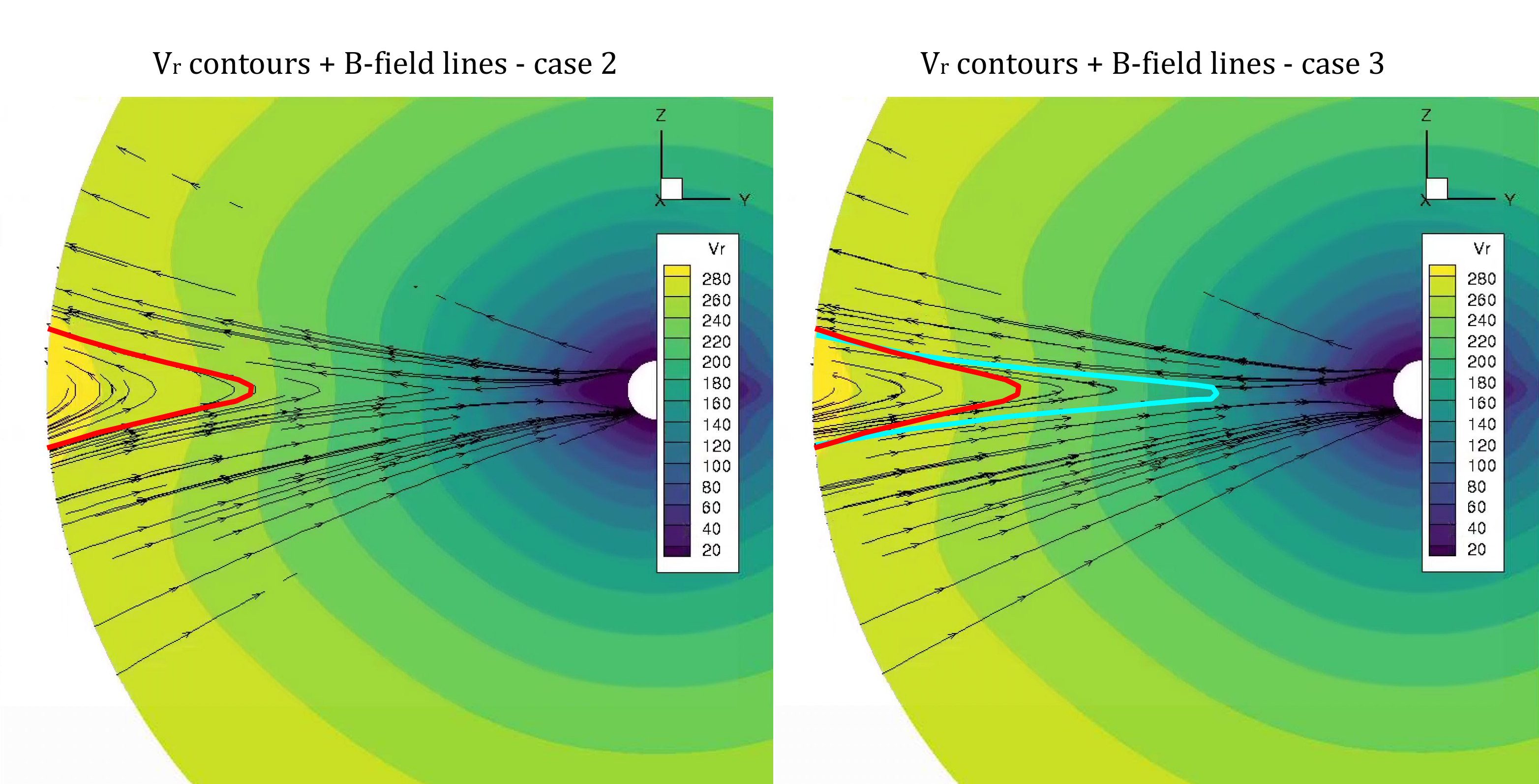}{0.9\textwidth}{}}
\caption{Comparison of the shape of the "ballooning" current sheet of case 2 and case 3 as a result of a different $B_r$ formulation. For easier comparison, the red line following a case 2 B-field line (left) is projected onto the case 3 flowfield (right), where the same line (with the same point where it starts and ends on the boundary) is followed in blue.}
\label{fig:ComparisonBallooningBr}
\end{figure*}

\textcolor{black}{The consequences this has on the flowfield and topology can be easily visualised when comparing the resulting magnetic field lines in Figure \ref{fig:ComparisonBallooningBr}. One of the field lines from case 2 is outlined in red and then projected onto the domain of case 3, where it is compared to a case 3 field line in blue that starts at the same point on the boundary. This illustrates that the field lines of case 3 are less divergent, especially close to the outer edge. }

\textcolor{black}{
The fact that the outer BC affects the magnetic topology so extensively even inside of the domain is concerning, especially since we cannot always ensure that there are no regions where the flow would expand super-radially (as will be shown later in case of a magnetogram-driven case).
}

\subsection{A BC-independent grid?}

\begin{figure*}[t!]
\centering
\gridline{\fig{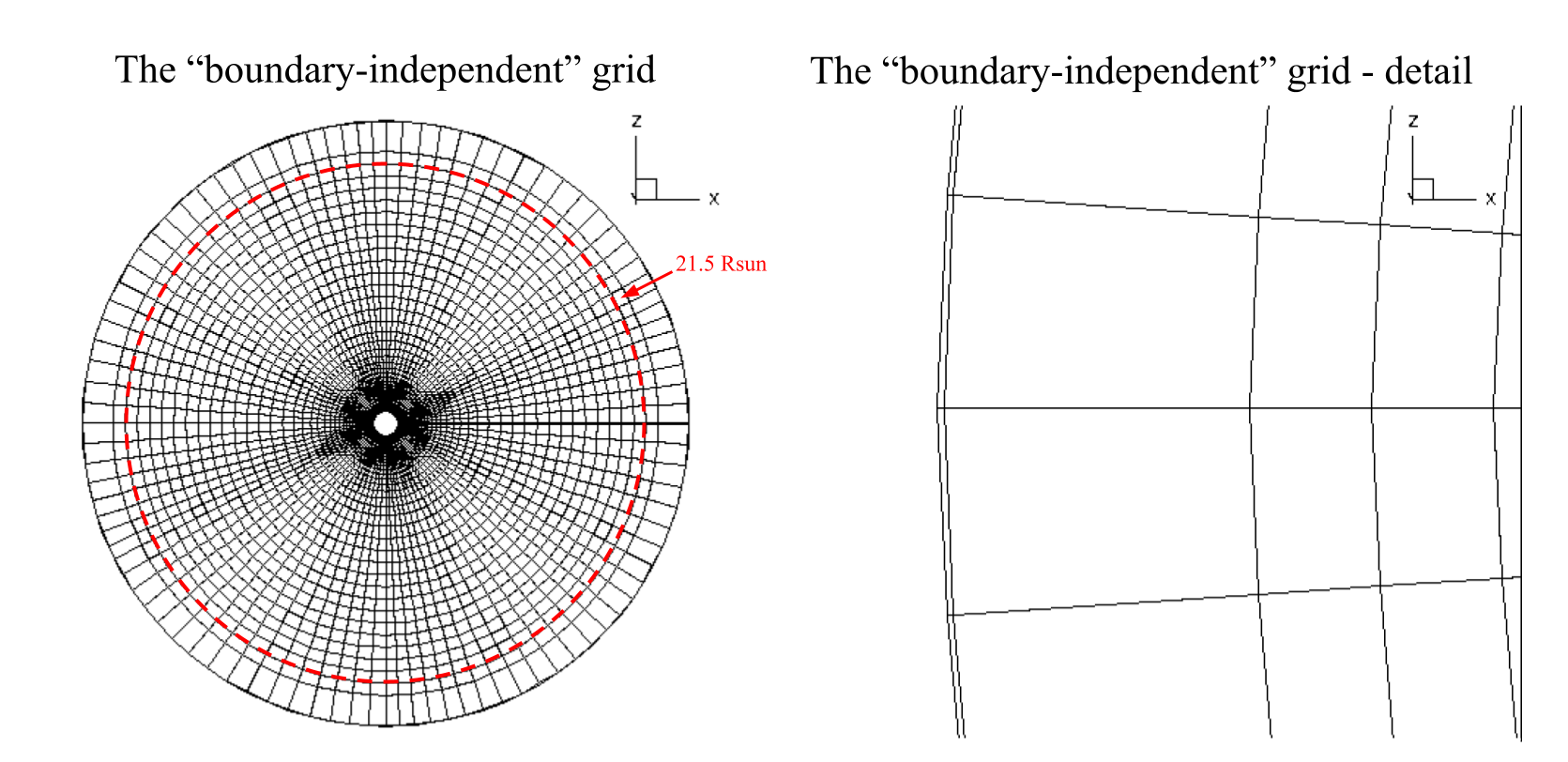}{0.9\textwidth}{}}
\caption{The "boundary-condition-independent" mesh developed for coupling with other codes. The ghost cell is very small, and thus the outer BC extrapolation errors are limited. In addition, the boundary is shifted further away from the 21.5$\;R_\odot$ from which the results are extracted, such that the solution at this coupling location is not affected by the BC formulation.}
\label{fig:NewMeshStructure}
\end{figure*}

\begin{figure*}[t!]
\centering
\gridline{\fig{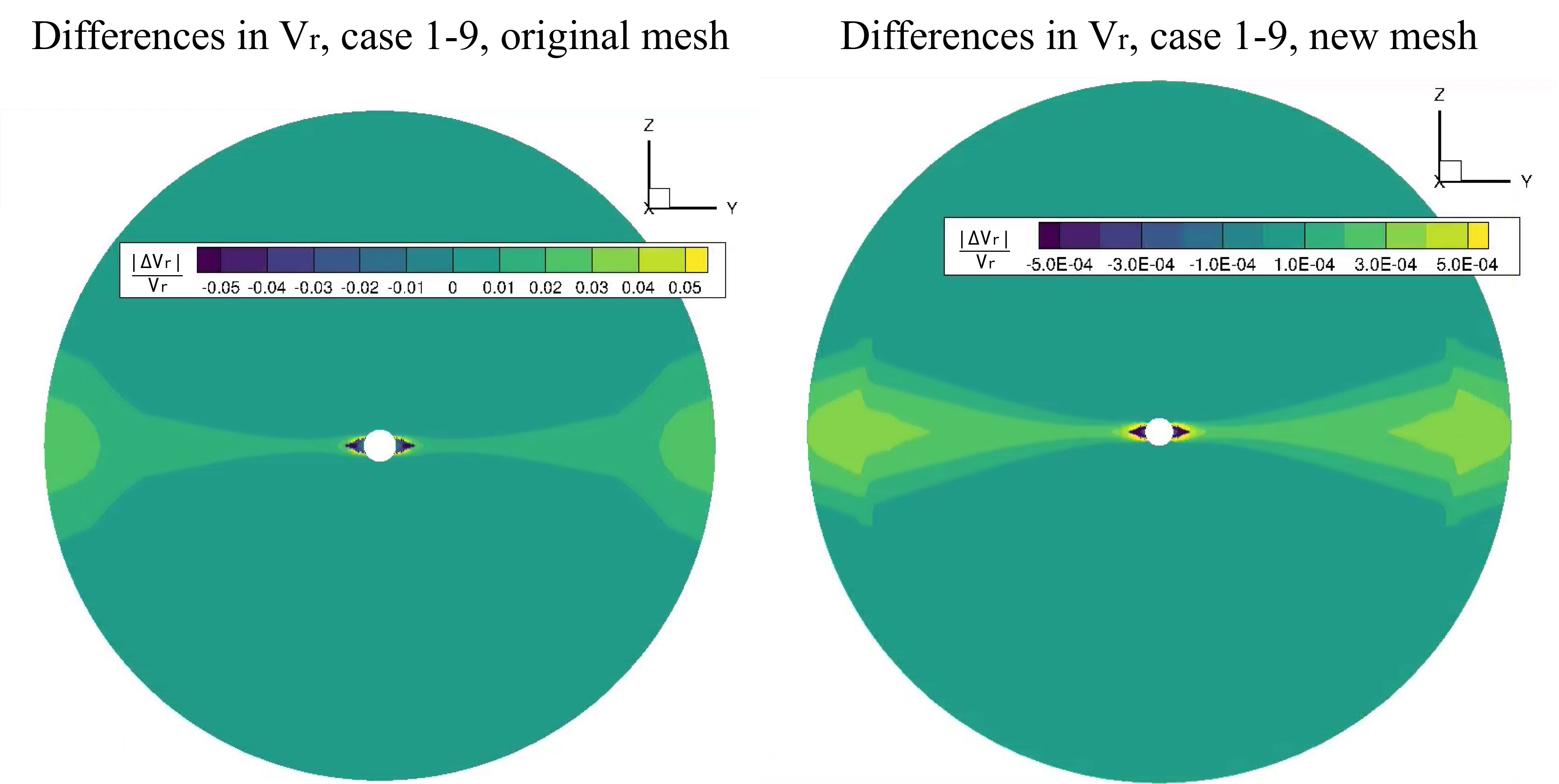}{0.99\textwidth}{}}
\caption{The relative difference in $V_r$ (in absolute sense) between case 1 and case 9, with the ordinary (left) and new (right) mesh. Before, the differences in the formulation of the BC between cases 1 and 9 led to relatively large differences in the velocity profile. With the improved mesh with smaller extrapolation effects, this is no longer the case. }
\label{fig:NewMesh}
\end{figure*}

\begin{figure*}[t!]
\centering
\gridline{\fig{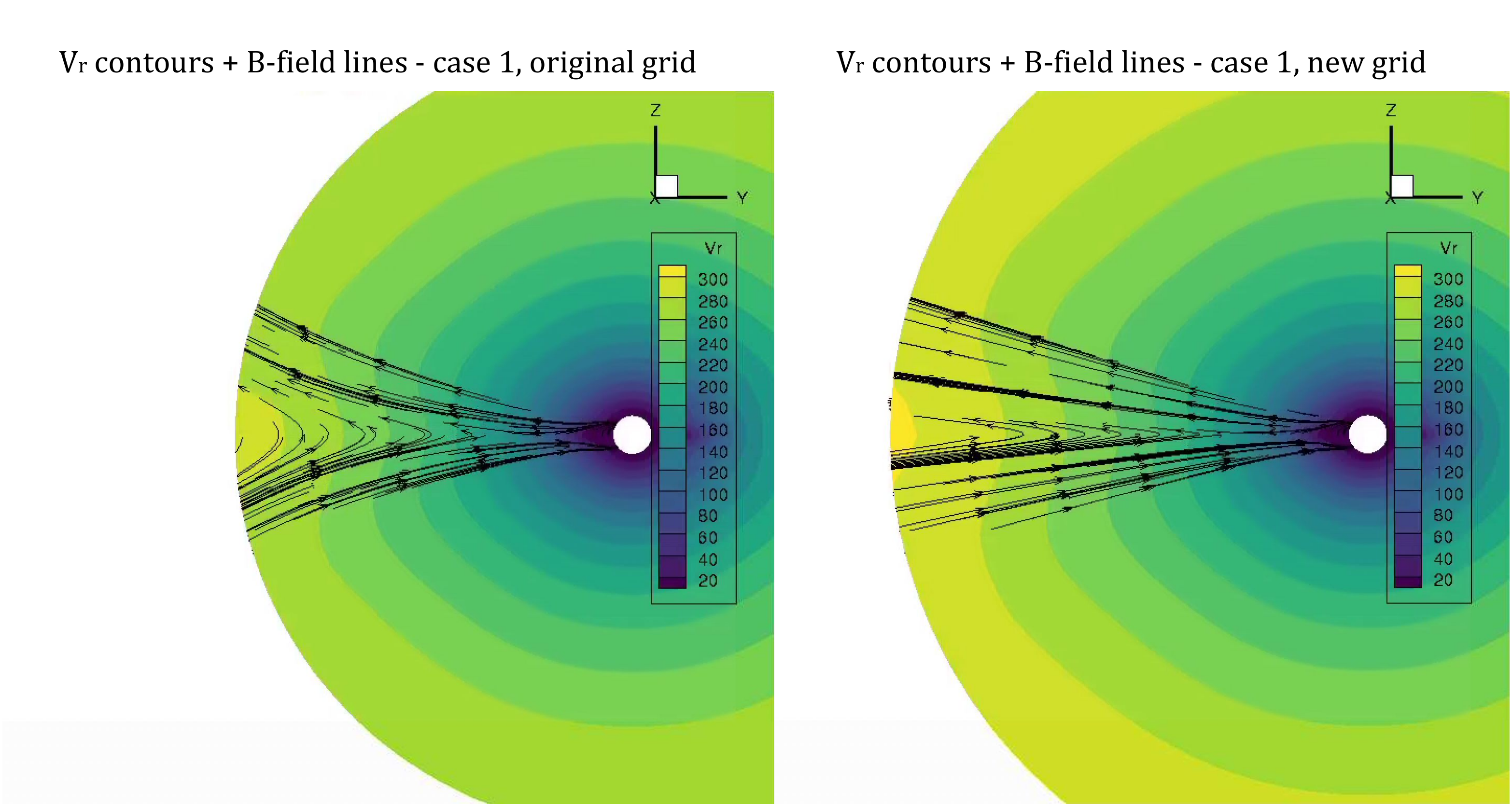}{0.9\textwidth}{}}
\caption{Demonstration of how the new grid design (on the right) with a small ghost cell can help prevent excessive opening of the streamers observed with the original grid (left) with the same outer BCs.}
\label{fig:ComparisonGridBallooning}
\end{figure*}

\textcolor{black}{As shown in the previous two sections, correct formulations of the BCs are not always straightforward and can effect the flow in the entire domain. Especially for cases where the BC formulations cannot be formulated universally, it would be useful to have additional tools to reduce the extrapolation error.}

The larger the region of extrapolation (the ghost cell), the larger the extrapolation errors are, which can then also more considerably spoil the solution in the rest of the domain. In the original grid which has been used so far, the last cell has the radial size of roughly 1$\;R_\odot$, meaning that the centre of the ghost cell is 0.5$\;R_\odot$ away from the boundary. If the last grid cell is made even larger, the effects become even more pronounced. Inverting this logic, with a very small edge cell, the ghost cell centre is also very close to the boundary and thus the extrapolation errors should be limited.

Moreover, the results on the outer boundary are very important from the perspective of coupling and transferring the CFD results to other software. Thus, not knowing how much of an effect the BC formulation can have on these results (at least locally), we do no longer deem it safe to apply these BCs exactly at the location from which the solution is to be read. It would be thus better practice to extent the domain beyond the coupling location of 21.5$\;R_\odot$ to make sure that the solution at the coupling location is not significantly affected in case the BC formulation is inadequate. 

\textcolor{black}{For this reason, an experimental grid was set up, which extends up to 25$\;R_\odot$ and which has a very small cell on the outer edge to demonstrate that the extrapolation error depends on the extrapolation distance}. With this new grid, cases 1 and 9 (i.e. the cases with arguably the largest differences between them in outer BC formulations) were re-run and compared. The new mesh is shown in Figure~\ref{fig:NewMeshStructure}, along with the 21.5$\;R_\odot$ location and the boundary detail.

\begin{figure*}[t!]
\centering
\gridline{\fig{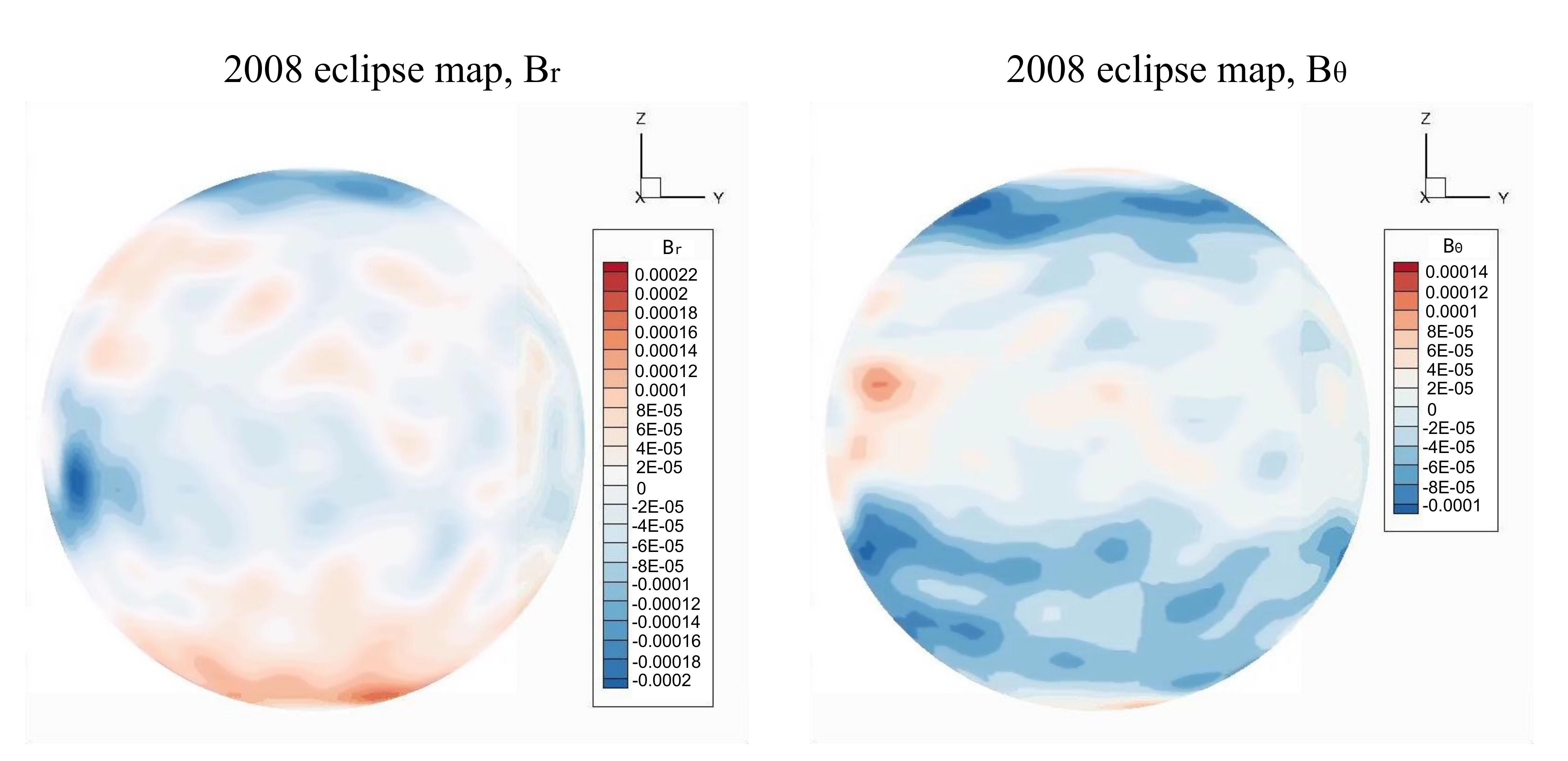}{0.9\textwidth}{}}
\caption{The magnetic field configuration, $B_r$ and $B_\theta$ from PFSS, of the 2008 eclipse.}
\label{fig:BrBtheta_map}
\end{figure*}

For both grids (the original and the new one), the relative differences (in absolute scale) in $V_r$ between the two cases were evaluated. These relative differences are plotted in Figure~\ref{fig:NewMesh}; on the left for the original grid and on the right for the new grid. Once again, we can see that on the former mesh with the large ghost cell (on the left), the outer BC formulation affects the solution even very close to the Sun. The differences with the new mesh are not even properly visible with the colour scale adjusted for the old grid; as they are 10 to 100 times smaller in magnitude everywhere in the domain. Clearly, with this new mesh design with a small ghost cell, the relative differences between the cases due to different extrapolation formulations are much lower. 

\textcolor{black}{It should be noted that this grid was designed in this way to merely demonstrate the effects of the size of the ghost cell. The fact that the last cell is so much smaller than the neighbouring cell, with such a large aspect ratio difference, could in practice cause problems with convergence. If that happens, a more gradual decrease in cell size near the outer boundary might be required at the cost of having to include more cells and thus increasing the computational demands. In our case however, with this grid, we have so far not observed any convergence problems even when resolving solar maxima, nor we have noticed any significant reduction in the computation speed.}


The above mentioned analysis of the inner and outer BCs considered a dipolar case only. To see how well the results of this analysis hold for a real case, the same analysis was repeated on a magnetogram-driven simulation. 


\section{Application to a Magnetic Map-Driven Simulation}
\label{sec:map}

\begin{figure*}[t!]
\centering
\gridline{\fig{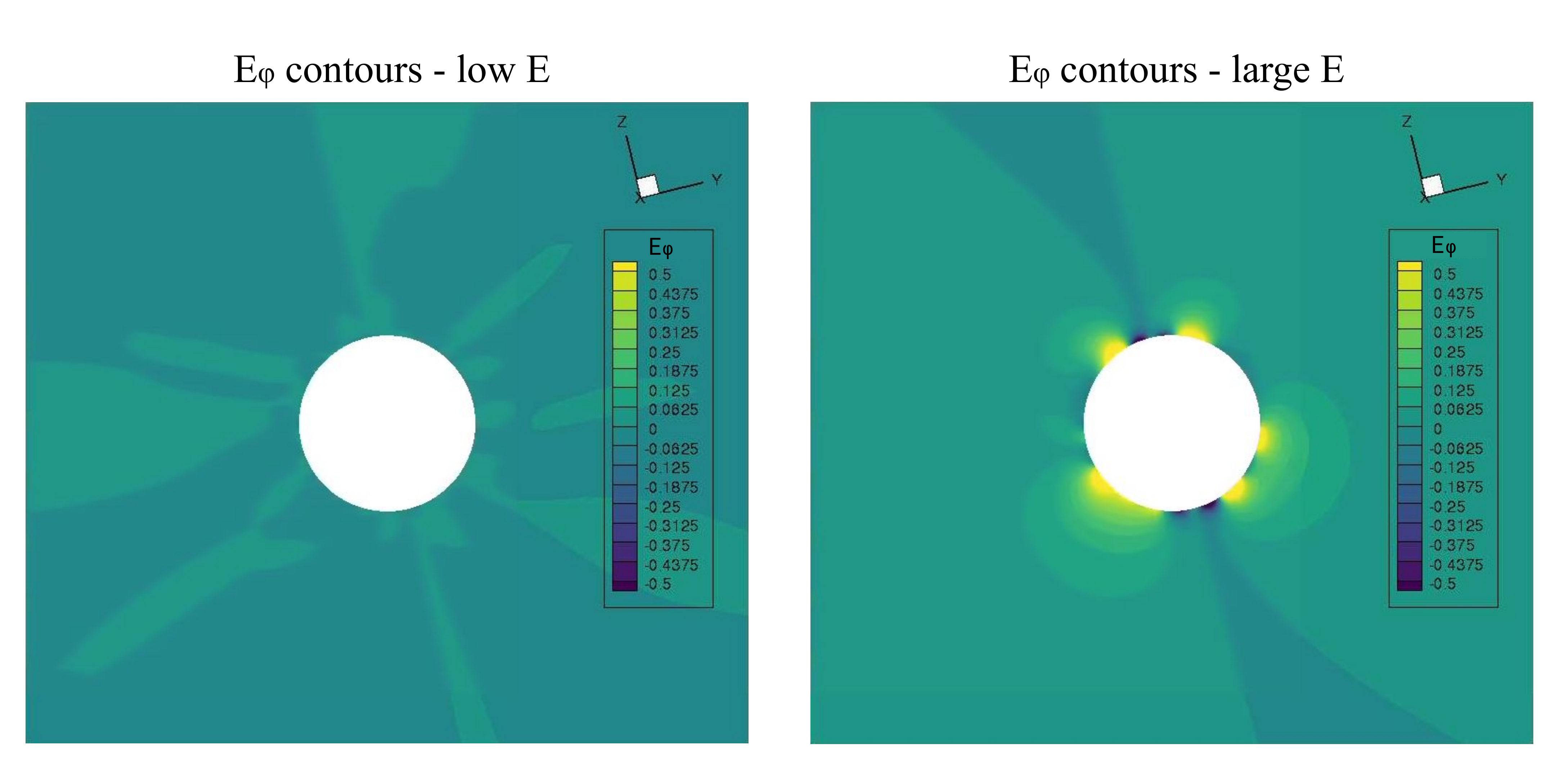}{0.9\textwidth}{}}
\caption{The $E_\varphi$ component for the two eclipse cases computed with and without the $\mathbf{E}$-field generating boundary.}
\label{fig:EphiZoom_map}
\end{figure*}

Having analysed BCs formulations on a simple dipole simulation, it is essential to also look at realistic magnetic map-driven simulations as these are the main intended applications for global coronal modelling. To this end, the solar eclipse from 2008 was investigated, the magnetic map of which (in combination with the PFSS solution for $B_\theta$) is shown in Figure~\ref{fig:BrBtheta_map}. The magnetic map used was an MDI map with $l_\text{max}$ of 25 \citep{Scherrer1995}. The projections used are the same as for the dipole, where a vertical cut (at y = 0 in the Cartesian system) was made. This projection is preferred unless observational comparisons must be made, since from the previous analysis of the dipole, in this configuration, it is already known where the mesh artifacts are located. 

First, let us observe the effects of the inner BCs and the artificial $\mathbf{E}$-field. Exactly as for the dipole, a "low $\mathbf{E}$" and a "large $\mathbf{E}$" simulations were run.

\begin{figure*}[t!]
\centering
\gridline{\fig{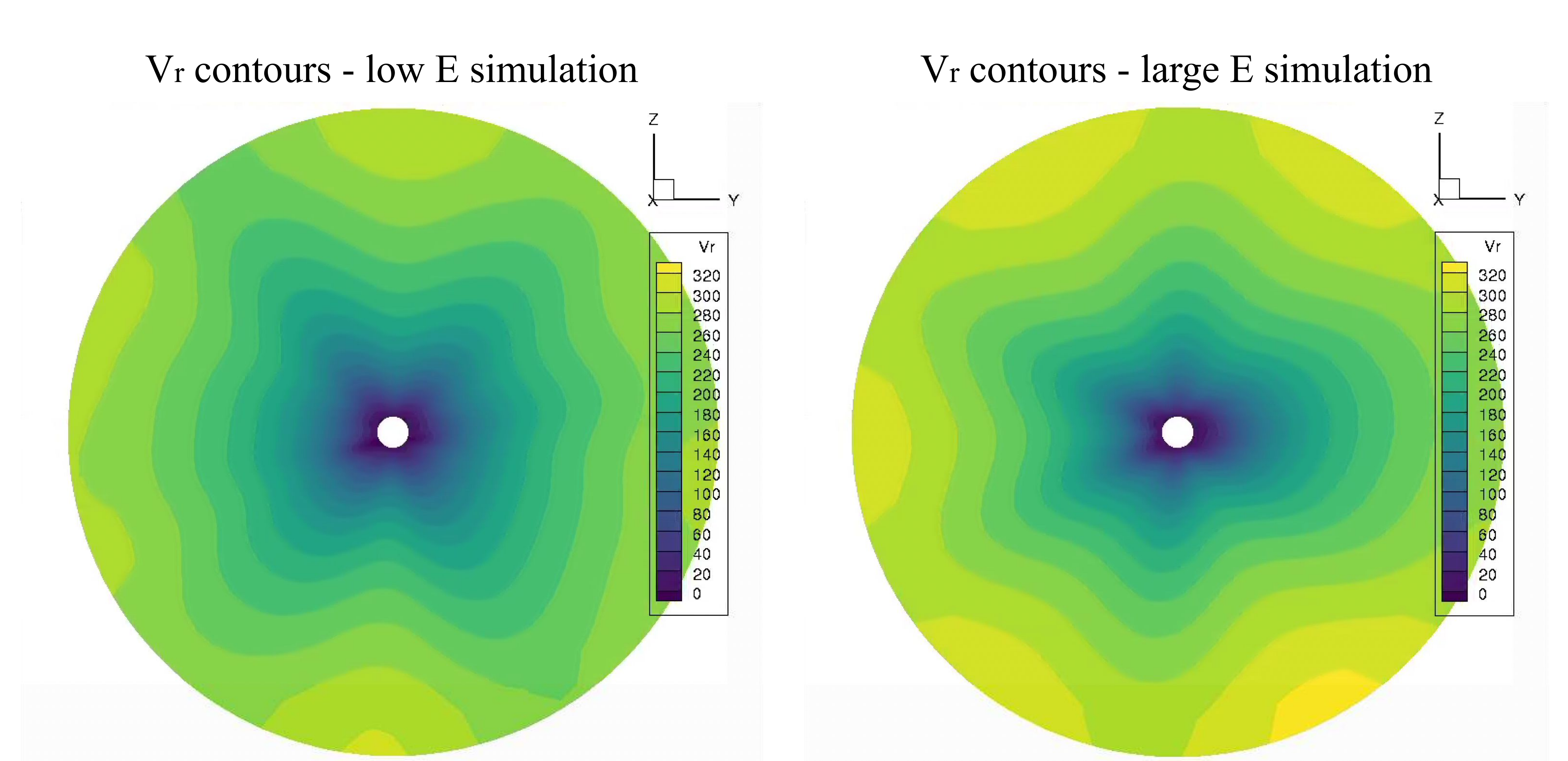}{0.9\textwidth}{}}
\caption{The $V_r$ contours for the "low $\mathbf{E}$" and "large $\mathbf{E}$" 2008 eclipse cases.}
\label{fig:Vr_map}
\end{figure*}

\begin{figure*}[t!]
\centering
\gridline{\fig{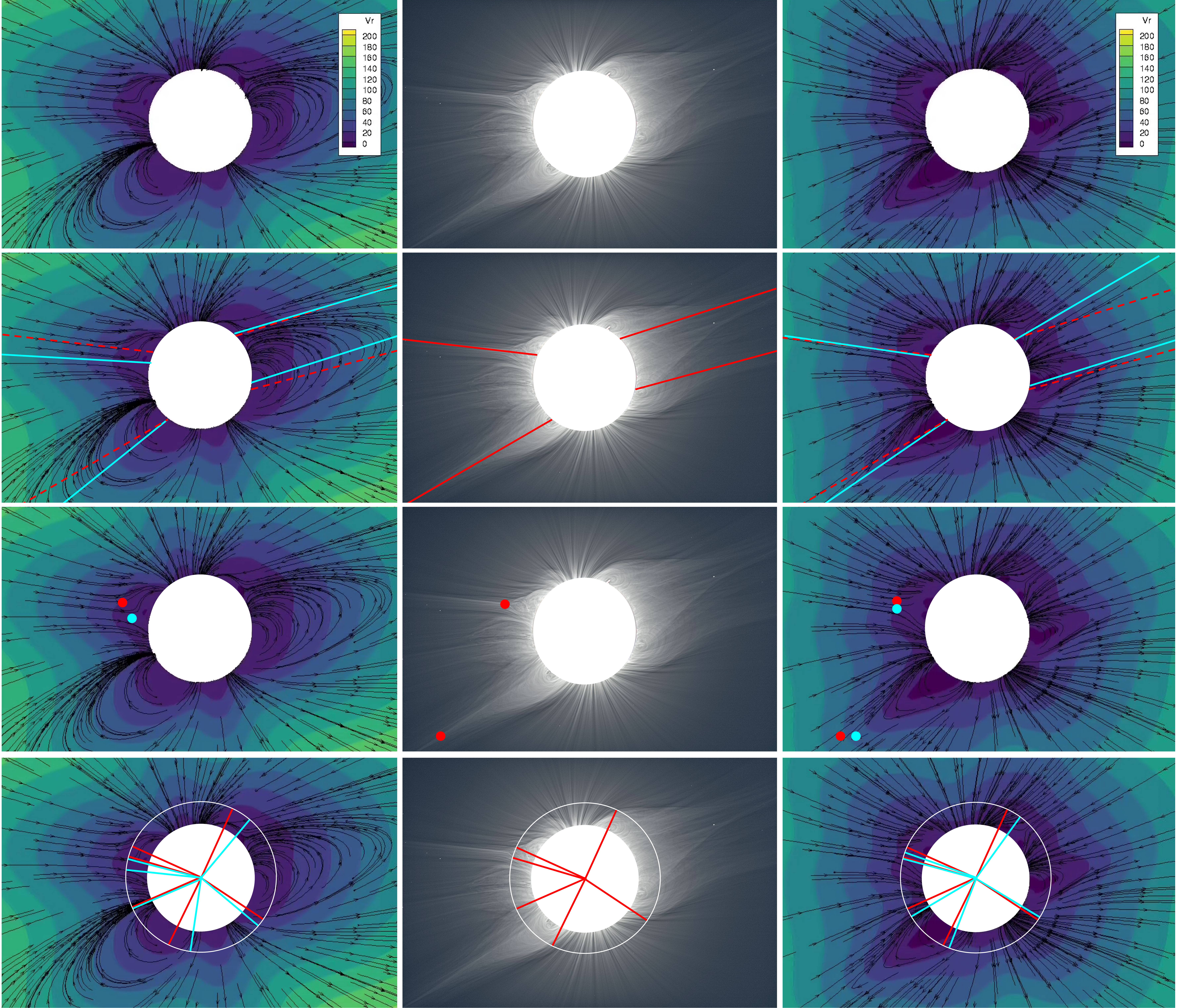}{0.9\textwidth}{}}
\caption{The comparison of the results of the 2008 simulation with "low $\mathbf{E}$" (right) and "large $\mathbf{E}$" (left) setups with the observations, using the framework by \cite{Wagner2022}. }
\label{fig:2008ComparisonObservationBlazejStyle}
\end{figure*}

Similarly to the dipolar results shown previously, these adjustments to the BC allowed the simulation to arrive to a result with much lower $\mathbf{E}$-field (locally at least 20 times lower everywhere in the domain) compared to the original "large $\mathbf{E}$" BC defined in Equations~(\ref{eq:poloidal1a}),  (\ref{eq:poloidal1b}) and (\ref{eq:poloidal2}), see Figure~\ref{fig:EphiZoom_map}.


These two results were compared to the 2008 eclipse observations by M.\ Druckmüller, P.\ Aniol and V.\ Rušin \footnote{\url{http://www.zam.fme.vutbr.cz}} in Figure~\ref{fig:2008ComparisonObservationBlazejStyle}. \textcolor{black}{In an ideal-case scenario, we could employ techniques such as white-light imaging to compare our results to the photos of the corona. However, since at this stage, we are still working with a polytropic model without any additional heating terms, the magnetic field is a much more reliable indicator of simulation accuracy compared to the density (which is required for white-light imaging).  To perform this comparison using the magnetic field with as much objectivity as possible, we adopt the approach of \cite{Wagner2022}, where we will focus on the resolution of streamers as the decisive factor. This is because of the fact that in the future, it is expected that these simulations will be also used to compute the evolution of flux ropes (which will be seeded on the inner boundary) for modelling of CMEs and as well as the propagation of the solar energetic particles. Both of these aspects are essential elements for space weather forecasting and both of these heavily depends on the resolution of the background coronal dynamics, such as streamers.}

\begin{figure*}[t!]
\centering
\gridline{\fig{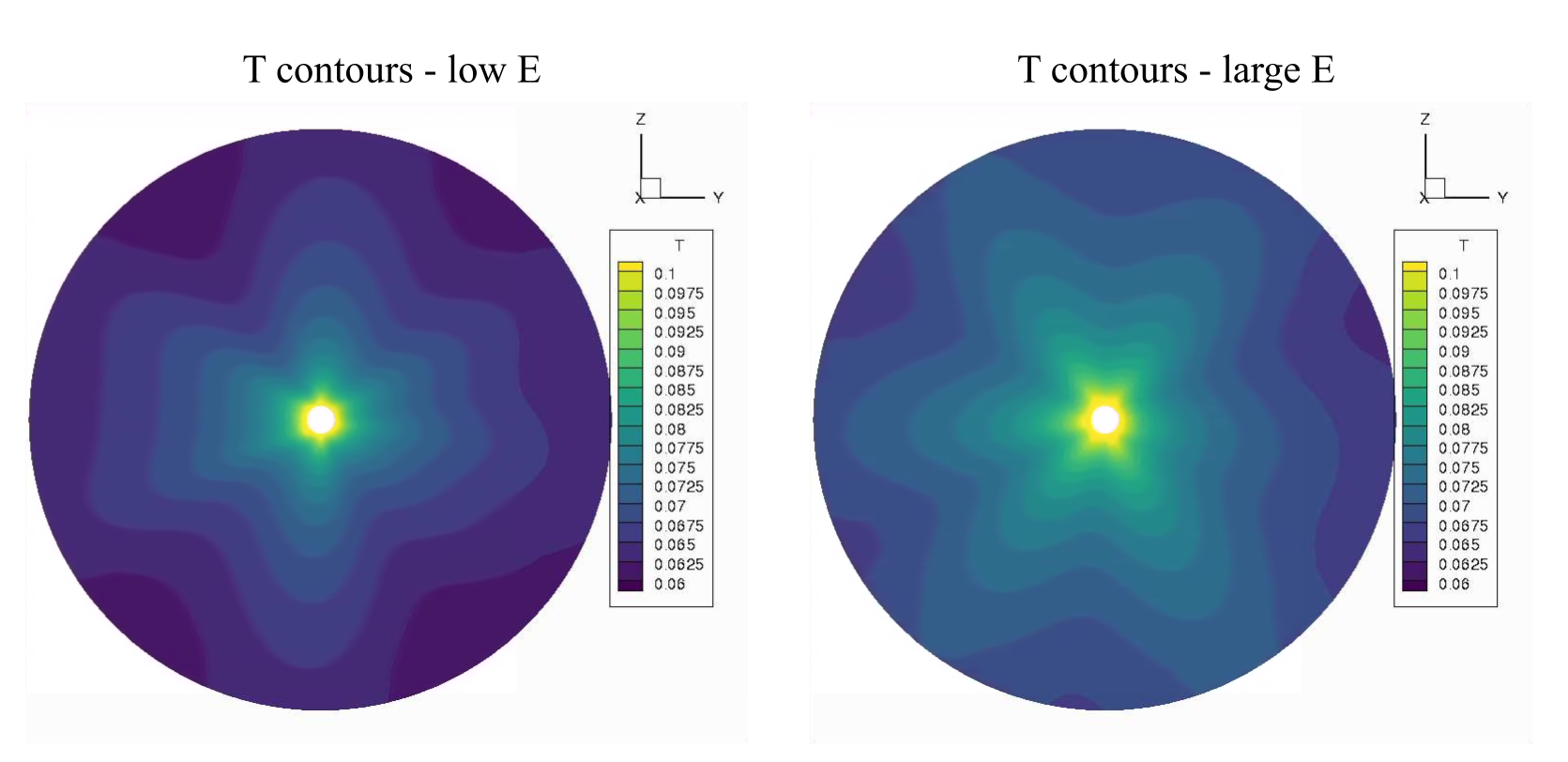}{0.9\textwidth}{}}
\caption{The comparison of the non-dimensional temperature profiles between the 2008 eclipse cases computed with (right) and without (left) an $\mathbf{E}$-field generating boundary.}
\label{fig:T_map}
\end{figure*}

\textcolor{black}{In Figure~ \ref{fig:2008ComparisonObservationBlazejStyle}, the first row shows the radial velocity contour in the background and the magnetic field lines plotted on the top of them, which were initiated randomly but on the same places for both simulations. The "low $\mathbf{E}$" simulation is on the right while the "large $\mathbf{E}$" on the left. In the second, third and fourth row of the Figure, we can see the comparison of the angle of the main streamer, the extent of the streamer and the size of the streamers, respectively. These aspects as determined from the observation are indicated in red and for each simulation result, the corresponding aspects are indicated in pale blue in the respective subfigures. The size of the streamers was determined at 1.3 $R_\odot$ - with an offset from the surface, just like in case of \cite{Wagner2022}. This offset was used due to the fact that the size of the streamers too close to the surface is mostly dependent on the magnetogram type and its resolution, while further away, the determination of the size from the observation becomes more and more difficult.}



\textcolor{black}{The biggest difference between the two models is observable already from the first row. It is the fact that in the case with the large electric field (on the left of the Figure), the $\mathbf{B}$-field lines are closed, round and do not copy well the observed sharp streamers. As a result, for example, the SE feature in case of the "large $\mathbf{E}$" seems to only consist of closed field lines, not forming a sheet surrounded by open field lines shown in the observations. The same holds for the dominant western feature. Apart from the upper western feature, the new BC can also reproduce the angle of the streamers better (second row). This result also better matches the extend of the streamers (row three) and slightly better the size of the features (row four). }

\begin{figure*}[t!]
\centering
\gridline{\fig{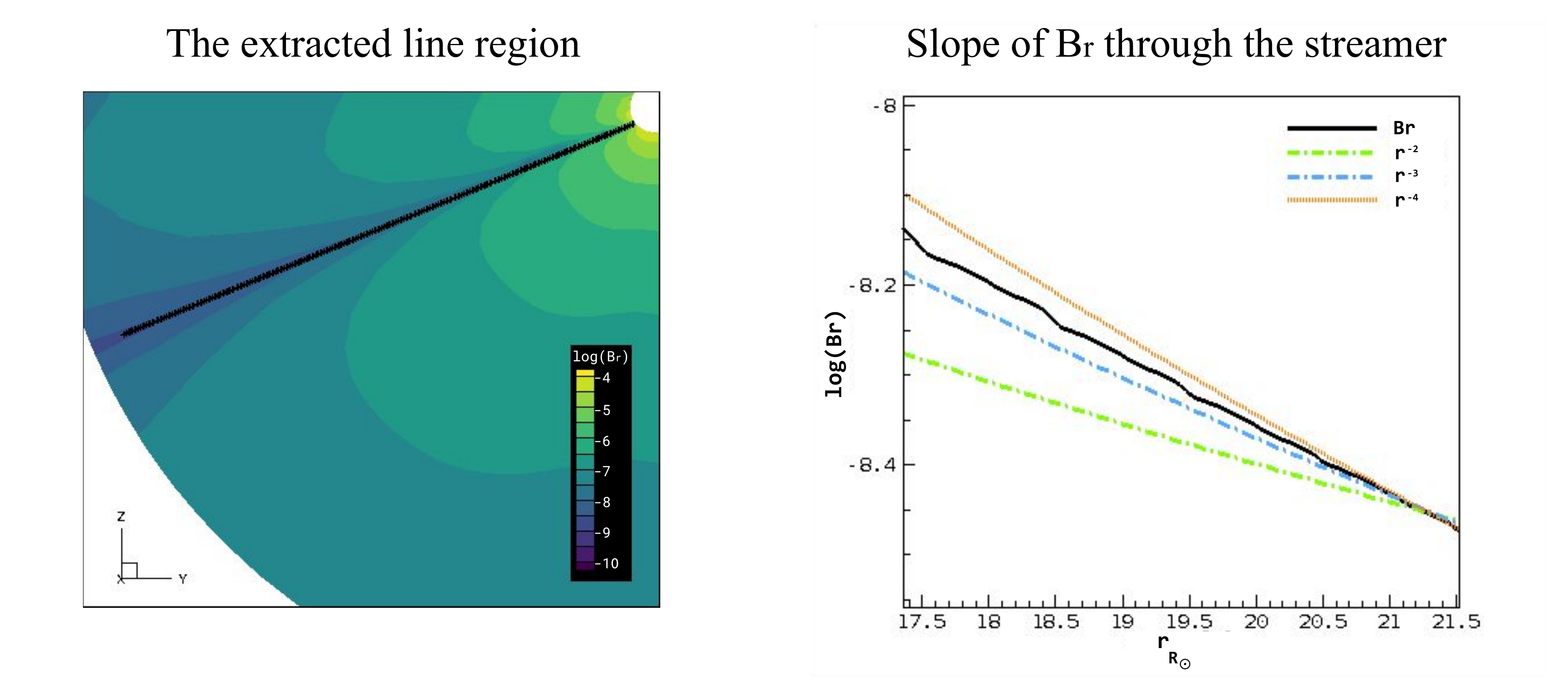}{0.99\textwidth}{}}
\caption{The $B_r$ slope (logarithmic, on the right) along a streamer from the 2008 eclipse simulation on the BC-independent, extended grid (shown as a black line on the left), showing a steeper grid than the expected $1/r^2$ scaling.}
\label{fig:Brnotalways}
\end{figure*}

Just like in the case of the dipole, while the differences in the sharpness of the features were significant, the most striking differences were seen in the temperature profile. The temperature profiles for the two cases, here defined simply as the ratio of the non-dimensional pressure and density, are shown in Figure~\ref{fig:T_map}. \textcolor{black}{The two cases have completely different temperature distributions and magnitudes, despite having the same temperature prescribed on the inner boundary. The fact that such important differences can arise as a result of the implicit prescription of the electric field, to the Authors' knowledge, has never been highlighted in existing literature. }
Next, since we could see in case of the dipole that the magnetic structure of the current sheet was heavily influenced by the extrapolation error, here we applied the same analysis to demonstrate these effects on real magnetogram. We have examined the $B_r$ profile along one such streamer; the lower left high-intensity streamer of the 2008 map. In this analysis, the corrected, extended, small-ghost cell grid was used for the map simulation and thus the slope at 21.5$\;R_\odot$ should not be heavily affected by the presence and the formulation of the outer BCs. The cut and the corresponding $B_r$ profile are shown in Figure~\ref{fig:Brnotalways}. Here, to determine the centre of the streamer along which the $B_r$ profile was extracted, the logarithm of the absolute value of $B_r$ is shown in the background of the contour plot. The location of the smallest $\log_{10}(B_r)$ corresponds to the location of the current sheet as was determined by the magnetic field lines.

\begin{figure*}[t!]
\centering
\gridline{\fig{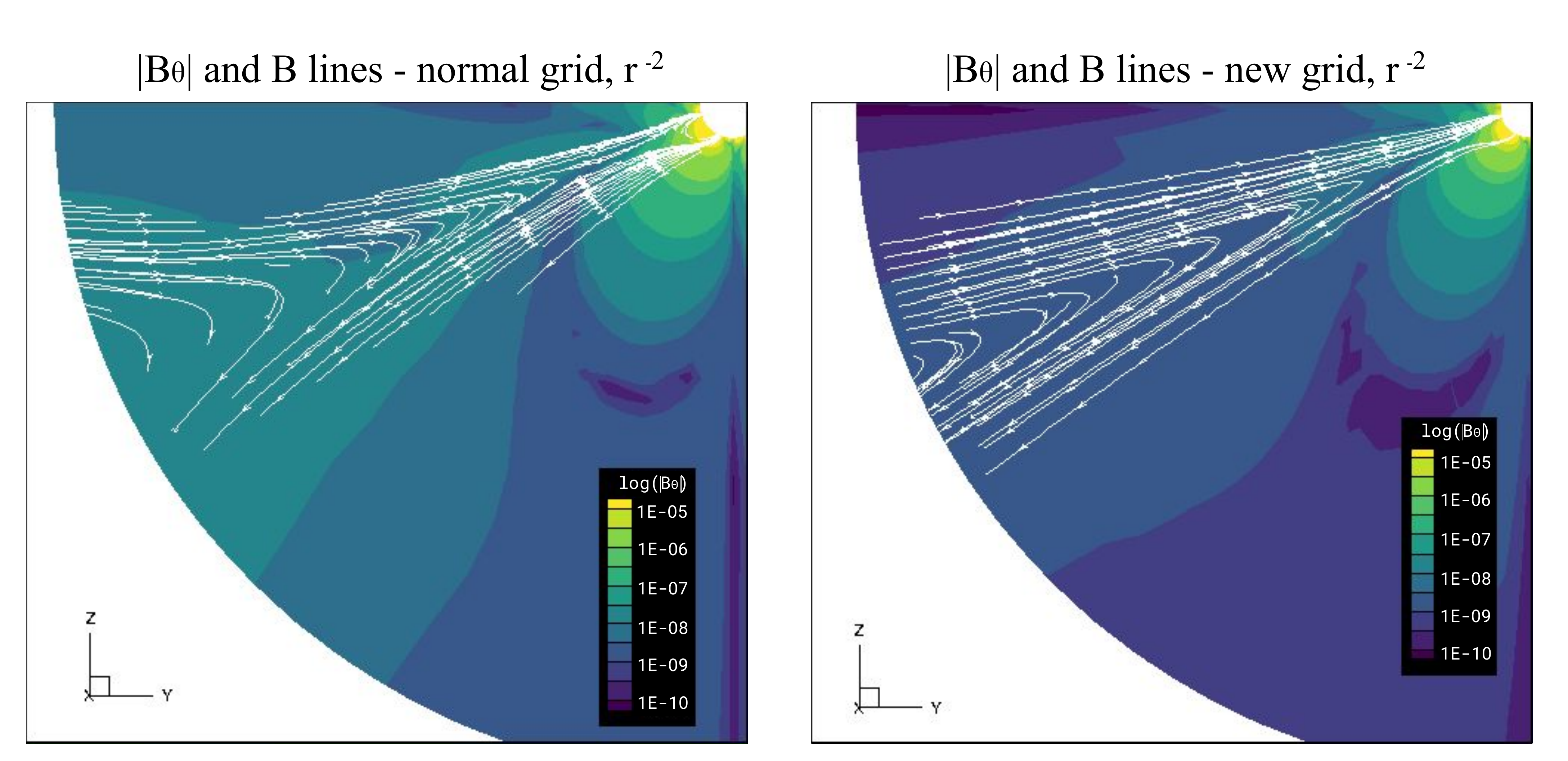}{0.99\textwidth}{}}
\caption{The comparison between the $\mathbf{B}$-field lines (with the absolute value of $B_\theta$ in the background) for the old and the new "BC-independent" grid, with both simulations extrapolating $B_r$ through the inverse square power. The results for the old grid with excessive divergence are shown on the left, while the results with the adjusted grid with small ghost cells are on the right.}
\label{fig:LinesMapOutletBr}
\end{figure*}

Figure~\ref{fig:Brnotalways} shows that along this streamer, $B_r$ does no longer follow the inverse squared profile, but decreases much steeper. Actually, the slope, as shown from the fitting curve, lies somewhere between the third and fourth inverse power. \textcolor{black}{In case of the dipole, prescribing the descent of $B_r$ with a smaller power resulted in the streamers opening up more than in case of an accurate prescription. In this case, we see the same effect, with the difference that here even the $1/r^2$ law is insufficiently steep. This is, however, only limited to the region of the streamer.}

Thus, having a grid which limits the effect that the extrapolation can have on the $\mathbf{B}$-field topology, such as the grid introduced in the previous section, is truly an advantage. To demonstrate the effects of the grid alone, the 2008 map results on an ordinary grid with the $r^{-2}$ extrapolation were compared to the 2008 map results for the improved grid with small ghost cells, also using the $r^{-2}$ extrapolation. The magnetic field lines (along with the absolute value of $B_\theta$) are shown in Figure~\ref{fig:LinesMapOutletBr}, where the most prominent, lower left streamer is shown.

It is clear that the extrapolation error with the original grid leads to excessive divergence of the streamlines near the boundary due to the insufficient extrapolation power. The same locally-incorrect BC with the improved grid, however, does not show this excessive divergence, as the angle of the $\mathbf{B}$-field lines does not change near the boundary.

In summary, from the magnetic map-based application, it can be seen that setting the initial state and the inner boundary to produce no electric field leads to quantitatively and qualitatively different solutions, with more physical temperature profiles, sharper streamers and a better reproduction of the observational data. Using a grid which limits the errors due to extrapolation on the outer boundary, on the other hand, prevents excessive divergence of the streamers without the need to locally modify the extrapolation laws.


\section{Conclusions and Recommendations}
\label{sec:conclusion}

\textcolor{black}{In this study, we have focused on numerical aspects of global coronal modelling. We started with the inner BC, to which we got through an analysis of the ideal-MHD formulation compared to the full multi-fluid MHD.} Due to the removal of the magnitude of the $\mathbf{E}$-field from the MHD equations, which in practice represents how well the magnetic and velocity structures are aligned and focused, the ideal-MHD equations can lead to qualitatively and quantitatively very different results simply depending on the initial state and BC formulations. It is thus important to monitor the $\mathbf{E}$-field in the simulations and aim to reduce the amount of it generated artificially. Two different techniques to avoid excessive $\mathbf{E}$-field were discussed in this paper; i)~ensuring that the simulation is initiated from a completely $\mathbf{E}$-field free state and ii)~adjusting the inner BC to prevent the generation of the $\mathbf{E}$-field on the inner surface.

\textcolor{black}{Here, the two types or simulation cases were compared; one with a large electric field in the domain and one} with a low electric field. It was shown that the artificially generated $\mathbf{E}$-field leads to streamers being resolved less accurately on the same grid. Furthermore, it was also found that the resulting temperature profile had a fundamentally different shape when the $\mathbf{E}$-field was minimised compared to the simulation with the spurious-$\mathbf{E}$-field-generating BC. The new temperature profile was following the structures of the streamers instead of the magnetically enhanced regions. 

\textcolor{black}{Briefly, we have also discussed the problem of additional reconnection in the domain due to numerical diffusion. We have shown that the location of the region where this happens depends on the refinement of the grid and the amount of magnetic field divergence in the domain.}

Next, also the outer BCs were discussed. Since many formulations are nowadays in use in which the variables are extrapolated to the ghost cells or onto the outer boundary, a parametric study was conducted to study the effects of these formulations in a systematic way. Nine different cases were compiled, where the BC formulations were varied for the velocity, magnetic field, pressure, density and the divergence cleaning parameter. Several conclusions were reached from this study; with some of the most notable being that:

\begin{itemize}
    \item the divergence cleaning parameter needs a zero Dirichlet outer BC for the simulation to converge;
    \item as long as the divergence cleaning parameter is set through a zero Dirichlet condition, the convergence performance of the other cases with varying outer BC formulations was almost the same;
    \item the aspects which seem to affect the solution the most are the extrapolation formulations of density and the radial $\mathbf{B}$-field component.
\end{itemize}

Afterwards, the extrapolation of $\rho$ and $B_r$ was investigated in more detail. This was done through running an extended-domain simulation up to 30$\;R_\odot$, such that the undisturbed profiles of these variables at the intended boundary, at 21.5$\;R_\odot$, could be studied. \textcolor{black}{It was found that at this distance, the density decreases roughly with $r^{-2.5}$ for our conditions and that the formulation of the $B_r$ outer BC can affect magnetic field topology in the entire rest of the domain - which can be problematic in case we locally obtain super-radially expanding flow.}
\textcolor{black}{To further limit outer-BC extrapolation errors beyond their careful formulation, an additional technique was presented; a new grid design}. This design has very small cells in the outermost radial layer, such that the ghost cells are also very small and very close to the boundary. In this fashion, the extrapolation errors due to the BC formulation are limited, as the magnitude of the extrapolation errors depends on the extrapolation distance. Secondly, this grid extends beyond 21.5$\;R_\odot$, which is the location from which the CFD solution should be transferred to another (heliospheric) code, to 25$\;R_\odot$. Thanks to this extension, even if there are still some effects due to inaccurate outer boundaries, these are not likely to affect the solution which is used as an input to other codes.

\textcolor{black}{Finally, all of the proposed adjustments were studied on the real magnetic map case of the 2008 eclipse, in order to allow for directly comparing our numerical results with observed coronal structures. Using} the new inner BC, the $\mathbf{E}$-field in the solution was found to be reduced by more than 20x, and \textcolor{black}{thus} allowed for a much sharper resolution of the coronal features on the same mesh. \textcolor{black}{The agreement between adjusted, sharper coronal features and observational data increased significantly.} The $B_r$ profile analysis was repeated, here not in a quiet region as before, but along one of the streamers in the solution on the larger grid with a small ghost cell. Along this streamer, the $B_r$ profile became much steeper compared to the quiet regions, and extrapolations with power-laws of $-3$ to $-4$ would locally have to be \textcolor{black}{applied}, as the extrapolation through the inverse square law led to excessive streamline divergence on the ordinary grid. This divergence was, however, prevented when the new grid-type was used instead.

Much work is still required to develop \textcolor{black}{a highly accurate and computationally efficient global coronal model which could work effortlessly on arbitrary real-time data. A number} of different models are currently in development globally, experimenting with a variety of numerical schemes, geometries, grids and MHD formulations, trying to \textcolor{black}{push forward the frontier of} current state-of-art. As we have noticed however, many times it is the small details in the numerical settings, such as a specific BC formulation, which can drastically improve the performance or the resolution of the model. For that reason, it is very important to keep paying attention to these small details and their effects, and to share these observations with the community if possible. In future work, we hope to further investigate the behaviour of ideal-MHD CFD for global coronal modelling from the numerical perspective, as for example focusing on topics such as determining the minimum required spatial resolution or the effects of using different types of magnetic maps and  map-smoothing functions. Once the solver is optimised to run with a reasonable accuracy and reliability, we aim to replace the steady-state magnetic maps with time-dependent simulations and flux rope injections. In addition, we also aim to develop a multi-fluid version which will resolve ions, electrons and neutrals as separate fluids. 

\vspace{0.5cm}
The authors would want to thank Fan Zhang for collaboration and constructive comments and insights. This work has been granted by the AFOSR basic research initiative project FA9550-18-1-0093. This project has also received funding from the European Union’s Horizon 2020 research and innovation programme under grant agreement No 870405 (EUHFORIA 2.0) and the ESA project "Heliospheric modelling techniques“ (Contract No. 4000133080/20/NL/CRS). These results were also obtained in the framework of the projects C14/19/089  (C1 project Internal Funds KU Leuven), G.0D07.19N  (FWO-Vlaanderen), SIDC Data Exploitation (ESA Prodex-12), and Belspo project B2/191/P1/SWiM. The resources and services used in this work were provided by the VSC (Flemish Supercomputer Centre),
funded by the Research Foundation - Flanders (FWO) and the Flemish Government. SOHO is a project of international cooperation between ESA and NASA.

\facility{Vlaams Supercomputing Centrum.}

\software{COOLFluiD \citep{Lani2005,Kimpe2005, Lani2013}, \url{https://github.com/andrealani/COOLFluiD/wiki}.}


\bibliography{main}{}
\bibliographystyle{aasjournal}

\end{document}